\documentclass[aps,prl,twocolumn,superscriptaddress,10pt,floatfix]{revtex4-2}

\usepackage{graphicx}
\usepackage{dcolumn}
\usepackage{bm}
\usepackage{amsmath}
\usepackage{amssymb}
\usepackage{braket}
\usepackage{nameref}
\usepackage{float}
\usepackage{placeins}
\usepackage{xcolor}
\usepackage[colorlinks=true,linkcolor=red,citecolor=blue,urlcolor=blue,]{hyperref}

\begin{document}

\title{Frequency Domain Berry Curvature Effect on Time Refraction}

\author{Shiyue Deng}
\affiliation{International Center for Quantum Design of Functional Materials (ICQD), Hefei National Research Center for Interdisciplinary Sciences at the Microscale, University of Science and Technology of China, Hefei, Anhui 230026, China}
\affiliation{Hefei National Laboratory, University of Science and Technology of China, Hefei 230088, China}

\author{Yang Gao}
\affiliation{International Center for Quantum Design of Functional Materials (ICQD), Hefei National Research Center for Interdisciplinary Sciences at the Microscale, University of Science and Technology of China, Hefei, Anhui 230026, China}
\affiliation{Hefei National Laboratory, University of Science and Technology of China, Hefei 230088, China}
\affiliation{School of Physics, University of Science and Technology of China, Hefei, Anhui, 230026, China}

\author{Qian Niu}
\affiliation{International Center for Quantum Design of Functional Materials (ICQD), Hefei National Research Center for Interdisciplinary Sciences at the Microscale, University of Science and Technology of China, Hefei, Anhui 230026, China}
\affiliation{School of Physics, University of Science and Technology of China, Hefei, Anhui, 230026, China}

\date{\today}
             
\begin{abstract}
We demonstrate that there exist frequency domain Berry curvature in the wave function of photons in dispersive optical systems. This property arises from the frequency dispersion of its dielectric function, which makes Maxwell equations a non-standard eigenvalue equation, with the eigenvalue (frequency) appearing inside the operator itself. We study this new Berry curvature effect on time refraction of magnetoplasmon-polariton as an example. It can induce deflection in the trajectory of a photon and make the ray swing.
\end{abstract}
\maketitle

Wave propagation in dispersive media, where the material response depends on frequency, leads to a non-standard eigenvalue problem, in which the frequency, serving as the eigenvalue, appears within the operator of the eigenvalue equation \cite{PhysRevB.50.16835,kuzmiak_photonic_1997,PhysRevLett.104.087401,Fietz:11,PhysRevLett.129.133903,PhysRevLett.132.126601,PhysRevA.111.042201}. Since the eigenvalue enters the equation in a nonlinear manner, such equations are also referred to as nonlinear eigenvalue equations \cite{PhysRevB.50.16835,PhysRevLett.132.126601,PhysRevA.111.042201}. We find in these systems there exists a novel geometric property --- frequency domain Berry curvature, which does not appear in standard systems such as Schrödinger equation \cite{RevModPhys.82.1959,PhysRevB.53.7010,PhysRevB.59.14915,PhysRevB.102.045423}, even in Floquet systems \cite{PhysRevB.106.224311}.

Maxwell equations in media are always non-standard because strictly speaking, all media are dispersive with respect to frequency. We should treat frequency as a parameter on an equal footing with momentum. For this reason, the electromagnetic wave has frequency domain Berry curvature in addition to momentum domain, which is ignored in previous studies \cite{PhysRevLett.93.083901,PhysRevLett.96.154802,PhysRevLett.100.013904,PhysRevA.78.033834} and should also be considered for a proper description of the propagation of the electromagnetic wave.  In this work, we develop a semiclassical theory based on a wave packet localized in spacetime to describe wave propagation under slow perturbations in non-standard eigenvalue systems, which additionally includes frequency domain Berry curvature. We apply the theory in magnetoplasmon system \cite{jackson1998classical,PhysRevLett.100.023902,jin_topological_2016,PhysRevLett.124.195001} and specifically study the dynamics of electromagnetic wave pulses under slow time modulation.

When a medium is modulated in time, the wave conserves its momentum but varies its frequency, this phenomenon is called time refraction \cite{morgenthaler_velocity_1958,stepanov_spectral_1968,mendonca_theory_2000,j_t_mendonca_time_2002,PhysRevA.62.033805,Khurgin:20,zhou_broadband_2020,lustig_time-refraction_2023,PhysRevLett.132.263802,dong_quantum_2024,PhysRevLett.134.073803}. Time refraction can be classified into two types, with the modulation being adiabatic \cite{stepanov_spectral_1968,Khurgin:20,zhou_broadband_2020, PhysRevLett.134.073803} and nonadiabatic \cite{PhysRevA.62.033805,lustig_time-refraction_2023,PhysRevLett.132.263802} ones, based on how much the refractive index changes over a single cycle. In this letter, we study the adiabatic time refraction of electromagnetic waves in a magnetoplasmon system. We find that, due to the non-zero Berry curvature in this system, the trajectory of an electromagnetic pulse becomes deflected and the deflection is a geometric quantity. As a consequence of the deflection, the ray swings. Our work provides a robust and convenient means to control light propagation via its geometric property.

\textit{\textbf{Theoretical framework}---}
We briefly introduce our semiclassical wave packet theory to describe wave propagation under perturbations in spacetime governed by non-standard eigenvalue equations. We first employ a method consistent with Refs.~\cite{PhysRevLett.132.126601,li2025geodynamicsartificialgravityspacetime,liu2025fullyfirstprinciplesapproachstudying}, introducing an auxiliary eigenvalue to standardize the eigenvalue equation. Subsequently, by superposing solutions of the auxiliary standard eigenvalue equation, we construct a wave packet localized in the eight-dimensional phase space (space-time and momentum-frequency), with the auxiliary eigenvalue at the wave packet center set to zero. Finally, we derive the equations of motion (EOM) of the wave packet by constructing its Lagrangian. The EOM incorporate Berry curvatures in all facets of the phase space. When the eigenvalue equation is standard, Berry curvature terms involving the frequency  all vanish, and the EOM coincide with those in Ref.~\cite{PhysRevB.59.14915}. Detailed procedures can be found in the supplemental material \cite{SM}.

\textit{\textbf{Model}---}
In this section, we give the specific model used in our study.
When considering only the photon degree of freedom, the Maxwell equations for dispersive materials represent an example of non-standard eigenvalue equation \cite{PhysRevLett.104.087401,Fietz:11,kuzmiak_photonic_1997,PhysRevLett.129.133903}. We consider a magnetoplasmon-ploariton model: a metallic medium placed in an external static magnetic field $\mathbf{B_0}=B_0\hat{z}$ \cite{jackson1998classical,PhysRevLett.100.023902,jin_topological_2016,PhysRevLett.124.195001}. We made lossless approximation in  dielectric tensor Eq.~\eqref{dielectric tensor}, which holds as long as the relaxation time
is much longer than the timescale of interest $(\sim
100\omega_{p}^{-1})$. Example systems include InSb \cite{WhalenWestgate1969,Cheeke1981,PhysRevResearch.2.023180} and
gas plasmas \cite{PhysRevLett.124.195001}. Neglecting dissipation and assuming that the radiational magnetic field is much smaller than the external magnetic field, its dielectric tensor can be described by the Drude model as following \cite{jackson1998classical,PhysRevLett.100.023902,PhysRevResearch.2.023180}:
\begin{equation}
    \label{dielectric tensor}
    \stackrel{\leftrightarrow}{\varepsilon}(\omega)=1-\frac{\omega_{p}^{2}}{\omega^{2}-\omega_{B}^{2}}\begin{pmatrix}1&-i\frac{\omega_B}{\omega}&0\\i\frac{\omega_B}{\omega}&1&0\\0&0&1-\frac{\omega_B^{2}}{\omega^{2}}\end{pmatrix},
\end{equation}
where $\omega_p=\sqrt{\frac{Ne^2}{m}}$ is the plasmon frequency of the metal,  $\omega_B=\frac{eB_0}m$ is cyclotron frequency. The external magnetic field breaks time-reversal symmetry. Next, we apply temporal modulation in this system, while spatial homogeneity is maintained. We let the plasmon frequency change over time \cite{PhysRevE.103.043207,wang_expanding_2025} in the manner of $\omega_{p}(t)=\omega_{p}(0)+\xi t$ with $\xi\ll\omega_{p}^{2}$ \cite{stepanov_spectral_1968}. Under such an adiabatic condition, the transmission of a single-mode light can be described by our wave packet theory. 

We now explain how our general framework is applied to this specific example. Maxwell's equations can be expressed in matrix form in frequency domain as \cite{PhysRevLett.100.013904,PhysRevA.78.033834,PhysRevA.81.053803},
\begin{equation}
\label{onshell_eq}
    \hat{\mathcal{H}}(\mathbf{k})|\psi(\omega,\mathbf{k},t_{c})\rangle=\omega\hat{\Theta}_{c}(\omega,t_{c})|\psi(\omega,\mathbf{k},t_{c})\rangle,
\end{equation}
where $\hat{\mathcal{H}}=\begin{pmatrix}
0 & -\mathbf{k} \times \\
\mathbf{k} \times & 0
\end{pmatrix}$, $|\psi\rangle=(\mathbf{E}, \mathbf{H})^T / \sqrt{2}$ is the electromagnetic wave function and $\hat{\Theta}_{c}(\omega,t_{c})=\begin{pmatrix}
\stackrel{\leftrightarrow}{\varepsilon}(\omega,t_{c}) & 0 \\
0 & \stackrel{\leftrightarrow}{\mu}
\end{pmatrix}$. We have used the local operator approximation in Eq.~\eqref{onshell_eq} as in Ref.~\cite{PhysRevB.59.14915}. The difference is that, in Eq.~\eqref{onshell_eq}, the eigenvalue, frequancy, appears in the operator $\hat{\Theta}_{c}$ of the eigenvalue equation, thus forming a non-standard eigenvalue equation. We have set the speed of light in vacuum $c=1$ in Eq.~\eqref{onshell_eq} and the permeability tensor $\stackrel{\leftrightarrow}{\mu}=1$ in our model. Eq.~\eqref{onshell_eq} is not a standard
eigenvalue equation. We
can standardize it by going “off-shell” and introducing an auxiliary eigenvalue $\lambda$:
\begin{equation}
\label{off-shell Maxwell eq}
\hat{\mathcal{L}}_{c}(\omega,\mathbf{k},t_{c})|\psi(\omega,\mathbf{k},t_{c})\rangle=\lambda_{c}\hat{\theta}_{c}(\omega,t_{c})|\psi(\omega,\mathbf{k},t_{c})\rangle,
\end{equation}
where $\hat{\mathcal{L}}_{c}=\hat{\mathcal{H}}-\omega\hat{\Theta}_{c}$ and $\hat{\theta}_{c}$ should be an artificially chosen (semi-)positive-definite operator. Here $\hat{\theta}_{c}(\omega,t_{c})=\text{diag}\{\theta,\theta,\theta,1,1,1\}$ is chosen as the energy density operator \cite{landau1995electrodynamics,jackson1998classical,PhysRevLett.100.013904} without the external magnetic field, where $\theta=1+\frac{\omega_{p}^{2}}{\omega^{2}}$. We call Eq.~\eqref{off-shell Maxwell eq} “off-shell equation" because only when $\lambda=0$ do the solutions of Eq.~\eqref{off-shell Maxwell eq} correspond to the real physical world. This standard eigenvalue equation for $\lambda$ will yield a set of orthonormal basis functions that satisfy the following conditions:
\begin{equation}
    \langle\psi_{m}(\omega,\mathbf{k},t_{c})| \hat{\theta}_{c}(\omega,t_{c}) | \psi _{n}(\omega,\mathbf{k},t_{c})\rangle=\delta_{m n},
\end{equation}
The inner product here includes integration over the spacetime coordinates $(\mathbf{r},t)$ and the subscripts “m" and “n" represent different bands. To describe the geometrodynamics of a particle, we construct a wave packet localized in spacetime, expressed as:
\begin{equation}
\begin{aligned}
    &\left|W\left(\omega_c, \boldsymbol{k}_c,t_c, \boldsymbol{r}_c\right)\right\rangle\\=&\int d \omega d \boldsymbol{k} a\left(\omega, \mathbf{k}\right)  \hat{\theta}^{\frac{1}{2}}_{c}\left(\omega, \boldsymbol{k}, t_c, \boldsymbol{r}_c\right)\left|\psi\left(\omega, \boldsymbol{k}, t_c, \boldsymbol{r}_c\right)\right\rangle.
\end{aligned}
\end{equation}
$a\left(\omega, \mathbf{k}\right)$ is a function localized around $\omega_{c}$ and $\mathbf{k}_{c}$ and the eigenvalue at the center of the wave packet satisfies $\lambda_{c}(\omega_{c},k_{c},t_{c})=0$. In the following, we omit the subscript “c" of the operator and eigenvalue. The selection of different $\hat{\theta}$ operators represents wave packets centered on different physical quantities. We chose $\hat{\theta}$ as “energy density operator at $B_{0}=0$" for mathematical convenience. For another physical quantity $\hat{\theta}^{\prime}$, its center position $\mathbf{r}^{{\theta}^{\prime}}_{c}$ has a dipolar correction from the previous one \cite{RevModPhys.82.1959,PhysRevLett.93.046602,PhysRevLett.124.066601}:
\begin{equation}
\label{dipolar correction}
\mathbf{r}^{{\theta}^{\prime}}_{c}=\frac{\langle \frac{1}{2}(\hat{r} \hat{\theta}^{\prime}+\hat{\theta}^{\prime}\hat{r})\rangle}{\langle\hat{\theta}^{\prime}\rangle}=\mathbf{r}^{\theta}_{c}+\frac{\left\langle\frac{1}{2}\left((\hat{r}-\mathbf{r}^{\theta}_{c})\hat{\theta}^{\prime}+\hat{\theta}^{\prime}(\hat{r}-\mathbf{r}^{\theta}_{c})\right)\right\rangle}{\langle\hat{\theta}^{\prime}\rangle},
\end{equation}
where $\langle \hat{\mathcal{O}} \rangle=\langle W|\hat{\theta}^{-\frac{1}{2}}\hat{\mathcal{O}}\hat{\theta}^{-\frac{1}{2}} |W\rangle$ for wave packet average, and $\mathbf{r}^{{\theta}}_{c}$ is the center of the previously chosen operator $\mathbf{r}^{{\theta}}_{c}=\langle\hat{\mathbf{r}}\rangle$.

Now we analyze the solutions of the eigenvalue equation. We focus on electromagnetic waves propagating in the plane perpendicular to the external magnetic field (x-y plane), now the eigen-modes of the electromagnetic waves can be decoupled into TE and TM ones due to the mirror symmetry perpendicular to the z-axis $\mathcal{M}_{z}$ \cite{joannopoulos2008photonic}. We consider the TE mode with $\psi_{TE}=\left(E_x,E_y,0,0,0,H_z\right)^{T}/\sqrt{2}$ (the electric field has components only in the x-y plane, whereas the magnetic field
has components only along the z-direction). Because of the breaking of time-reversal symmetry, the degeneracy of the TE and TM modes is lifted. For the TE modes, its dispersion relation is
\begin{equation}
\label{onshell_condition}
    k=\sqrt{\frac{\omega^{4} + \omega_{p}^{4} - \omega^{2} \left( \omega_{B}^{2} + 2\omega_{p}^{2} \right)}{\omega^{2} - \omega_{B}^{2} - \omega_{p}^{2}}},
\end{equation}
where $k=\sqrt{k_x^2+k_y^2}$.
From the off-shell eigen-equation \eqref{off-shell Maxwell eq}, we can solve for the dispersion $\lambda_{TE}$ and the eigen-state  $\psi_{TE}$ corresponding to the TE modes \cite{SM}. We then refer to Eq.~\eqref{onshell_condition} as TE on-shell condition.

Then we apply the EOM of the wave packet from the general framework to this system and they simplify to :
\begin{subequations}
\label{simple_EOM}
\begin{align}
\dot{\mathbf{r}}&=\mathbf{v}_{g}-(\Omega_{\mathbf{k}t}+\mathbf{v}_{g}\Omega_{\omega t}+\Omega_{\mathbf{k}\omega}\dot{\omega});\label{rdot}\\
\dot{\omega}&=P-\dot{\omega}\Omega_{\omega t};\label{omegadot}\\
\dot{\mathbf{k}}&=0.\label{kdot}
\end{align}
\end{subequations}
Hereafter, the subscript c denoting the center of the wave packet is omitted. $\mathbf{v}_{g}\equiv\frac{\partial \omega}{\partial \mathbf{k}}=\partial_\mathbf{k}\lambda/(-\partial_\omega\lambda)$ is group velocity, $P\equiv\frac{\partial \omega}{\partial t}=\partial_t\lambda/(-\partial_\omega\lambda)$ is named chirp rate \cite{PhysRevA.110.033519}. $\Omega_{p q}$ are Berry curvatures calculated at the center of wave packet defined as $\Omega_{p q}\equiv -2\operatorname{Im}\langle \partial_{p}v|\partial_{q}v\rangle$, where $|v\rangle=\hat{\theta}^{\frac{1}{2}}|u\rangle$ with $|u\rangle$ the amplitude part of the wave function $|\psi\rangle$ that satisfies normalization condition $\langle u | \hat{\theta} |u\rangle=1$, and $p,q=\{t,k_{x},k_{y},\omega\}$. The eigenvalue $\lambda$ and the wave function used in calculating Eq.~\eqref{simple_EOM} are functions of $\omega$, $\mathbf{k}$ and $t$. Ultimately, these functions must be restricted by the TE on-shell condition Eq.~\eqref{onshell_condition}. Eq.~\eqref{kdot} represents momentum conservation in this spatially homogeneous system, while Eq.~\eqref{rdot} and Eq.~\eqref{omegadot} represent adiabatic time refraction on the trajectory and frequency of an electromagnetic pulse, including corrections from the Berry curvature. Using Eq.~\eqref{simple_EOM}, one can determine the trajectory and frequency at any moment of an electromagnetic pulse propagating in this system. Notably, the “energy" $\lambda$ of the wave packet is conserved in Eq.~\eqref{simple_EOM}. If the wave packet is initially on-shell (satisfy Eq.~\eqref{onshell_condition}), it remains on-shell throughout its evolution in phase space.

In the following, we illustrate the physics of time refraction described by Eq.~\eqref{rdot}. In this isotropic system, the group velocity of a wave packet $\mathbf{v}_{g}$ is parallel to the momentum $\mathbf{k}$. The temporal modulation $\xi$ is suddenly applied at $t=0$ and the velocity acquires an anomalous component $\mathbf{v}_{a}=-(\Omega_{\mathbf{k}t}+\mathbf{v}_{g}\Omega_{\omega t}+\Omega_{\mathbf{k}\omega}\dot{\omega})$ that is perpendicular to $\mathbf{v}_{g}$. If we regard the frequency as an auxiliary variable, one sees that when differentiating with respect to any other parameter, one must also include its indirect dependence on the frequency. By defining an on-shell Berry curvature $\Omega_{\mathbf{k}t}^{on}=\Omega_{\mathbf{k}t}+\frac{\partial \omega}{\partial \mathbf{k}}\Omega_{\omega t}+\Omega_{\mathbf{k}\omega}\frac{\partial \omega}{\partial t}$, the anomalous velocity above reduces to the familiar Thouless-pumping term \cite{PhysRevB.27.6083}. The group velocity remains parallel to the momentum direction through the process while the magnitudes of the group velocity and the anomalous velocity change during the time evolution. We show this process in Fig.~\ref{velocity}.
\begin{figure}[!htbp]
    \label{velocity}
    \centering
    \includegraphics[width=\columnwidth]{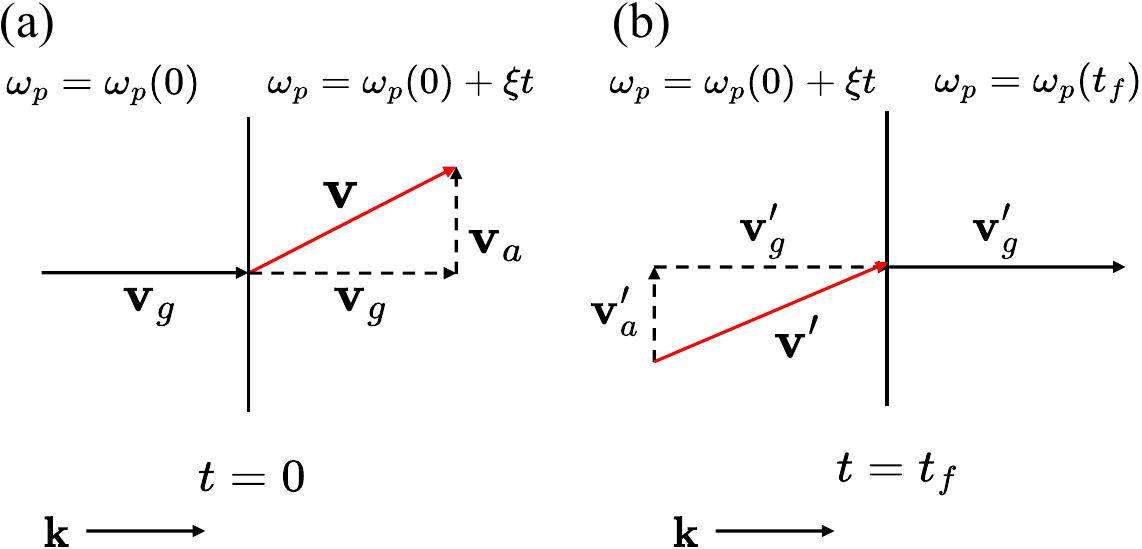}
    \caption{(a) The temporal modulation $\xi$ is applied at $t=0$ and the pulse's velocity acquires an anomalous component that is perpendicular to the group velocity. (b) $\xi$ is removed at $t=t_{f}$ and the pulse loses the anomalous velocity immediately.} 
    \label{velocity}
\end{figure}

\begin{figure}[!htbp]
    \centering
    \includegraphics[width=\columnwidth]{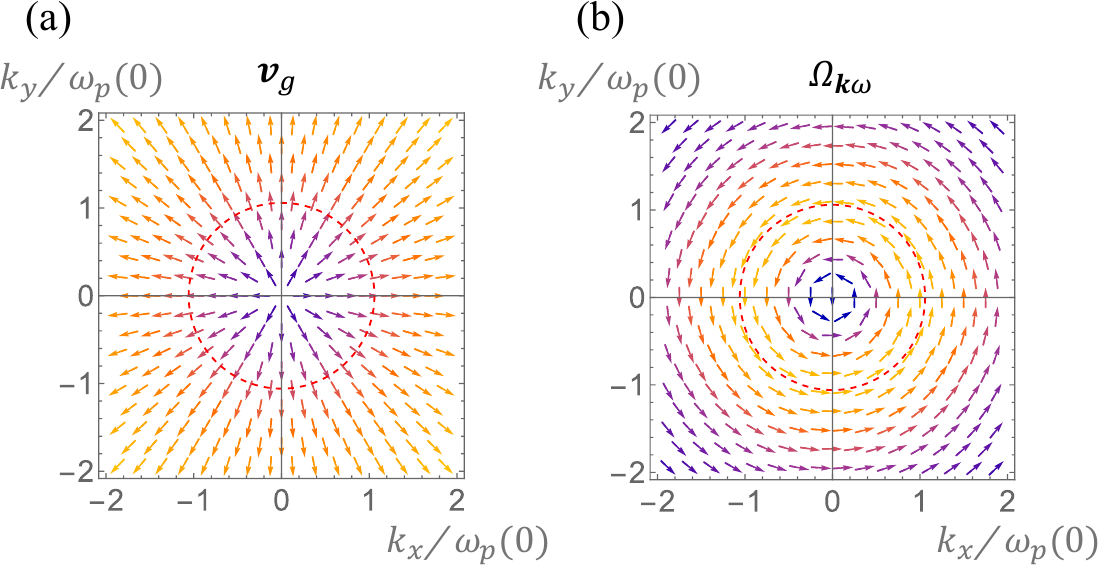}
    \caption{(a) Group velocity $\mathbf{v}_{g}$. (b) Berry curvature $\Omega_{\mathbf{k} \omega}=(\Omega_{k_{y} \omega},\Omega_{k_{x} \omega})$ in $k_{x}-k_{y}$ plane at $t=0$ with parameters $\omega_{B}=\omega_{p}(0)$, $\xi=0.01\omega_{p}^{2}(0)$ and $\omega=1.75\omega_{p}(0)$. The red dashed circles represent the on-shell condition $k(\omega=1.75\omega_{p}(0))=1.059\omega_{p}(0)/c$.}
    \label{in_plane_fields}
\end{figure}

\begin{figure}[!htbp]
    \centering
    \includegraphics[width=\columnwidth]{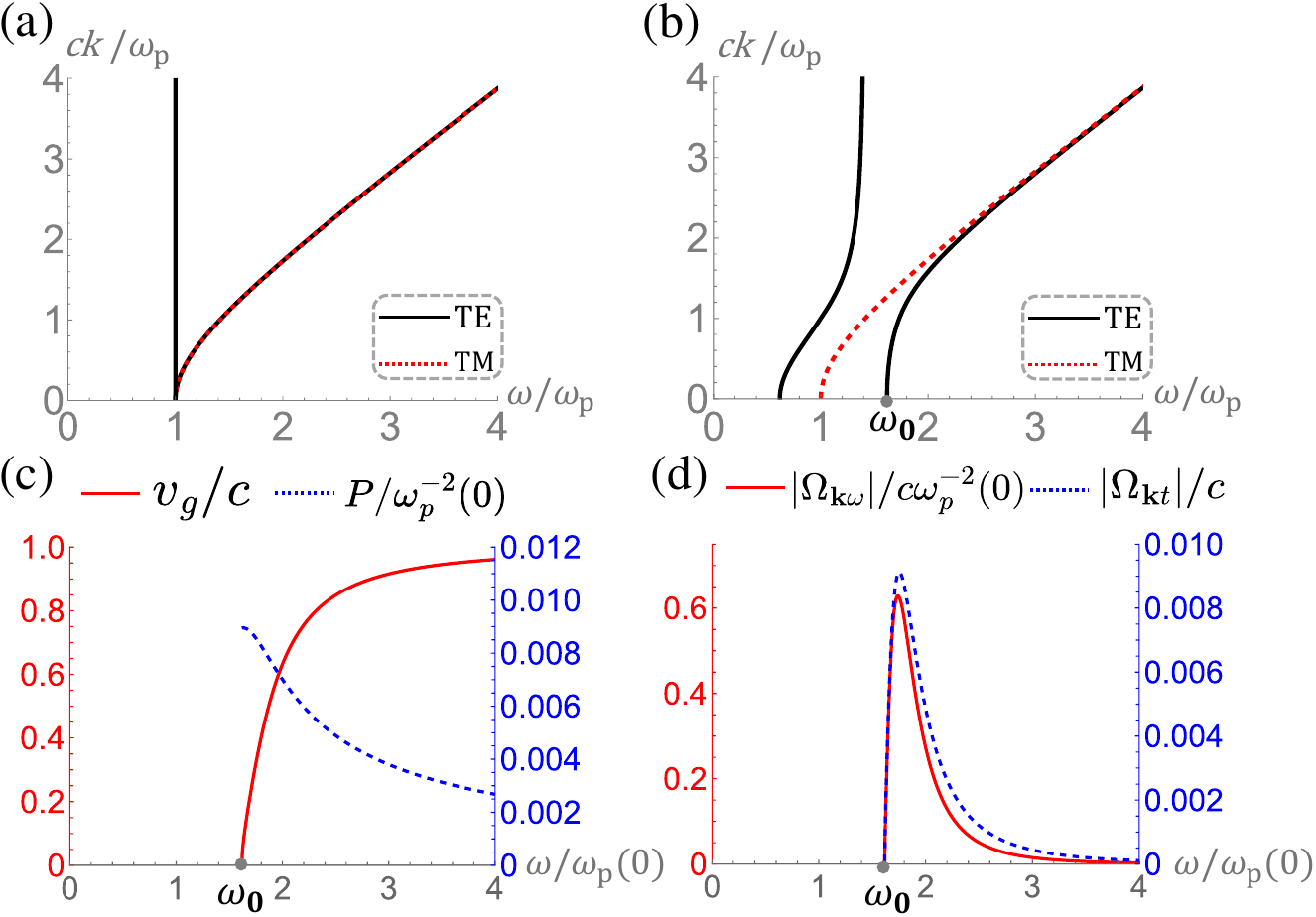}
    \caption{Dispersion curves (a) without external magnetic field and (b) with finite external magnetic field $\omega_{B}=\omega_{p}$. (c) The group velocity $\mathbf{v}_{g}$ and chirp rate $P$. (d) Magnitudes of  Berry curvatures $\Omega_{\mathbf{k} \omega}$ and $\Omega_{\mathbf{k} t}$ of the upper branch TE mode calculated under the on-shell condition \eqref{onshell_condition} as a single-valued function of $\omega$ at $t=0$ with $\xi=0.01\omega^{2}_{p}(0)$.}
    \label{onshell_emf}
\end{figure}

\textit{\textbf{Time-dependent Berry curvature fields}---}
In this part, we analyze the Berry curvature field involved in this context.
Fig.~\ref{in_plane_fields} shows $\mathbf{v}_{g}$ and $\Omega_{\omega \mathbf{k}}$ involved in Eq.~\eqref{simple_EOM} for the upper TE band. Other terms are not shown in the picture because $P$ is a scalar field that has a relatively simple structure, while $\Omega_{\mathbf{k} t}$ has the same distribution and direction as $\Omega_{\omega \mathbf{k}}$ and $\Omega_{\omega t}=0$ in this case \cite{footnote}. The distribution of these fields maintains the rotational symmetry of the system. Moreover, $\Omega_{\mathbf{k} \omega}$ and $\Omega_{\mathbf{k}t}$ are everywhere perpendicular to $\mathbf{v}_{g}$ as described by Fig.~\ref{velocity}. Therefore, $\Omega_{\mathbf{k} \omega}$ and $\Omega_{\mathbf{k}t}$ lead to a new type of Hall effect of light (unlike the spatially inhomogeneous case discussed in Ref.~\cite{PhysRevLett.93.083901}). All terms in Eq.~\eqref{simple_EOM} need to be calculated under the on-shell condition Eq.~\eqref{onshell_condition}, in which $\omega$ depends only on the magnitude of $\mathbf{k}$. In Fig.~\ref{in_plane_fields}, this condition is expressed as the red dashed circle.
Then we investigate Berry curvatures under the on-shell condition Eq.~\eqref{onshell_condition}, which are single-valued functions of $\omega$, as we present in Fig.~\ref{onshell_emf} along with the dispersion relation defined by Eq.~\eqref{onshell_condition}. Figs.~\ref{onshell_emf}(a) and~\ref{onshell_emf}(b) show the dispersion curves in the absence of $\mathbf{B_{0}}$ and for $\omega_{B}=1$, respectively. In our dispersion plots, the horizontal and vertical axes are opposite to those commonly used in the previous literature to make it easier to compare with the Berry curvature plotted as a function of frequency. When $\mathbf{B_{0}}=0$, one branch of the TE mode has a constant frequency $\omega_{p}$ (longitudinal), and the other branch is degenerate with the TM mode (transverse) \cite{PhysRevLett.134.183801}. Once $\mathbf{B_{0}}$ is turned on, the TM dispersion remains unaffected, but the two branches of the TE mode hybridize with each other and open a gap, lifting the degeneracy between the TE and TM modes. The hybridization leads to nonzero Berry curvatures which are zero at $k=0$, corresponding to the frequency $\omega_{0}=1.618\omega_{p}$ (the golden ratio) and reach a peak and then gradually decrease back to zero as the frequency increases. The group velocity gradually increases and approaches the speed of light in vacuum, while the chirp rate decreases with frequency. The two parts of the anomalous velocity,
$\Omega_{\mathbf{k} \omega} P$ and $\Omega_{\mathbf{k} t}$, are of the same order of magnitude, both around $10^{-3}c$ at maximum in this setting.

\begin{figure}[!htbp]
    \centering
    \includegraphics[width=\columnwidth]{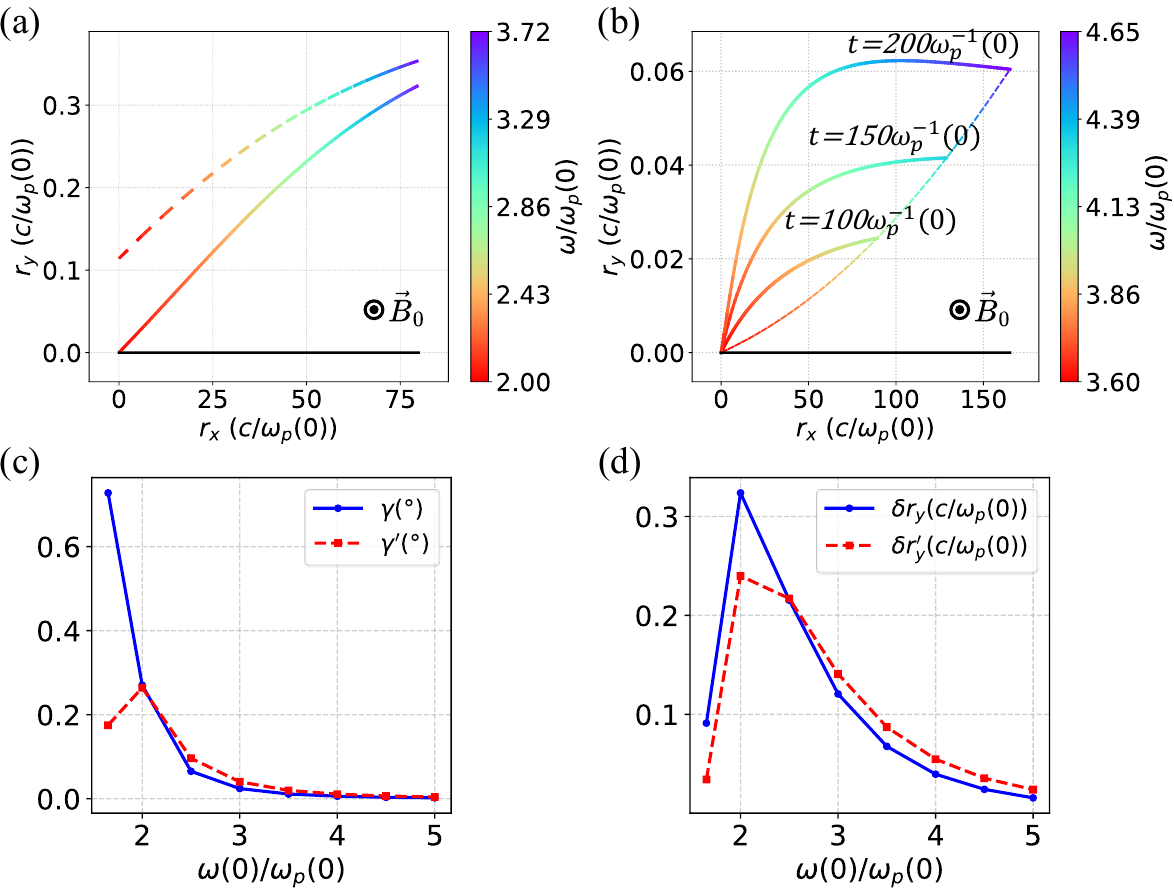}
    \caption{(a) The trajectories of $\hat{\theta}$ center (solid line) and $\hat{\theta}^{\prime}$ center (dashed line) for a pulse with initial frequency $\omega(0)=2\omega_{p}(0)$ emitted at $t = 0$. (b) Ray swings (solid lines) for continuously emitted pulses with the same initial frequency of $\omega(0) = 3.6\omega_{p}(0)$ The dashed line represents the trajectory of the first emitted pulse. In both (a) and (b), the color maps the frequency of the pulse at the moment it reaches that point, and the black rays denote the pulse's trajectory in the absence of the time modulation. The variations of (c) the initial deflection angle and (d) the total displacement with the initial frequency, where 
    $\gamma$ and $\delta r_{y}$ corresponds to the center of $\hat{\theta}$, and 
    $\gamma^{\prime}$ and $\delta r_{y}^{\prime}$ represent the center of $\hat{\theta}^{\prime}$.
}
    \label{trajectory}
\end{figure}

\textit{\textbf{Trajectories and ray swings}}---
Finally, we study the dynamics in real space of an electromagnetic wave packet or a light bullet \cite{PhysRevLett.134.073803}. Given an initial pulse frequency $\omega(0)$, we compute the corresponding on-shell momentum from Eq.~\eqref{onshell_condition}, and then use Eq.~\eqref{simple_EOM} to calculate the evolution of its position and frequency. The solid line of {Fig.~\ref{trajectory}(a) depicts the trajectory of a pulse with an initial frequency $\omega(0)=2\omega_{p}(0)$}, where the color of a point on the trajectory represents the central frequency of the pulse at the moment it reaches that point. The momentum is set to be along the x-axis. Due to the sudden transition of the terms $\Omega_{\mathbf{k}t}$ and $\dot{\omega}$ in Eq.~\eqref{rdot} from zero to finite values at the moment $t=0$, the direction of the pulse velocity undergoes an instantaneous finite change, as illustrated in Fig.~\ref{velocity}(a). During this whole process, the displacement of the pulse perpendicular to its momentum direction is given by the integral of the anomalous velocity over time:
\begin{equation}
\label{displacement}
\begin{aligned}
    &\delta \mathbf{r}_{y} =\int_{0}^{t_{f}} \mathbf{v}_{a}(t) dt\\
    =&\int_{\omega_{p(0)}}^{\omega_{p}(t_{f})}\frac{(\partial_{\mathbf{k}}\lambda)\Omega_{\omega \omega_{p}}+(\partial_{\omega_{p}}\lambda)\Omega_{\mathbf{k} \omega}-(\partial_{\omega}\lambda)\Omega_{\mathbf{k} \omega_{p}}}{\partial_\omega\lambda} d\omega_{p}.
\end{aligned}
\end{equation}
where we have used the relations $\partial_{t}=\dot{\omega}_{p}\partial_{\omega_{p}}$ and $d\omega_{p}=\dot{\omega}_{p}dt$. We also replaced $\dot{\omega}$ with $P$ in the first-order approximation. From Eq.~\eqref{displacement}, one can see that for a pulse with certain momentum, the displacement of its trajectory perpendicular to the momentum direction depends only on the initial and final values of the parameter $\omega_{p}$, and not on how fast it changes, as long as the adiabatic approximation $\dot{\omega}_{p}\ll \omega_{p}^{2}$ holds.
Based on the trajectory of the center of $\hat{\theta}$, we can use Eq.~\eqref{dipolar correction} to calculate the trajectory of the center of any other physical quantity. For example, for the total energy center $\hat{\theta}^{\prime}=\partial_{\omega}(\omega\hat{\Theta})$ \cite{landau1995electrodynamics,jackson1998classical,PhysRevLett.100.013904}, Eq.~\eqref{dipolar correction} can be simplified as \cite{SM}
\begin{equation}
    \mathbf{r}^{{\theta}^{\prime}}_{c}-\mathbf{r}^{\theta}_{c}=\frac{\operatorname{Im}\langle \partial_{\omega}u|\hat{\mathcal{L}}|\partial_{\mathbf{k}}u\rangle}{\partial_{\omega}\lambda}.
\end{equation}
The dashed line in Fig.~\ref{trajectory}(a) shows the trajectory of $\hat{\theta}^{\prime}$ center,
Fig.~\ref{trajectory}(c) and (d) present the variations of the initial deflection angle and total displacement of centers of energies with and without magnetic field with the initial frequency, respectively. The initial deflection angle becomes large when $\omega$ is close to $\omega_{0}$, because the x-component of velocity, i.e., the group velocity, is relatively small there. The trend of total displacement is similar to the Berry curvatures in Fig.~\ref{onshell_emf}(d), since they both are directly related to the anomalous velocity. The total displacement is $\sim10^{-1}c/\omega_{p}$ at the frequency near the band bottom $\omega_{0}$, which is about $10\mu m$ for InSb \cite{PhysRevResearch.2.023180,buddhiraju_absence_2020} and $1 mm$ for gaseous plasma \cite{PhysRevLett.124.195001}.

Next, we consider the case of continuously emitted pulses, as shown in Fig.~\ref{trajectory}(b). In this case, pulses emitted at different moments will experience different environments, resulting in distinct motion trajectories. We assume that these pulses are incoherent. Suppose that we begin emitting pulses at $t=0$ and observe later, the shape of light will be given by the connecting lines of the current positions of all previously emitted pulses. The region where energy is transmitted is the finite area formed by the motion trajectories of all the pulses. Starting from $t=0$, pulses with a frequency of $\omega(0)=3.6\omega_{p}(0)$ and $\mathbf{k}=k\hat{x}$ are continuously emitted from the origin. We plot the trajectory of the pulse emitted at $t=0$ as the dashed line. Then, for a fixed observation time $t>0$, we connect the points corresponding to that moment on all the trajectories of the previously emitted pulses, with the connecting lines represented by solid lines, which represent the shape of light at that moment. The appearance of this line changes over time, which is the ray swinging phenomenon we mentioned previously. The region enclosed by the solid and dashed lines represents the area in which energy is transmitted at the corresponding moment $t$. This process is shown in the supplemental animation.

We point out that the Berry curvature fields in this context satisfy two source-free Maxwell equations. The Berry curvature tensor satisfies the Bianchi identity $\partial_{\lambda}\Omega_{\mu\nu}+\partial_{\mu}\Omega_{\nu\lambda}+\partial_{\nu}\Omega_{\lambda\mu}=0,\quad\mu,\nu,\lambda=0,1,2,3$. In our model, the parameter space $\{t,k_{x},k_{y},\omega\}$ is four-dimensional, similar to the 3+1 dimensional real space $x^{\mu}=\{t,x,y,z\}$ and reciprocal space $k^{\mu}=\{\omega$,$k_{x}$,$k_{y}$,$k_{z}\}$. Thus, we can define gauge fields in a manner analogous to the electromagnetic field strength tensor $F_{\mu \nu}$, which satisfies Faraday's law $\nabla\times\mathbf{e}+\partial_{t}\mathbf{b}=0$ and Gauss's law for magnetism $\nabla\cdot\mathbf{b}=0$, where $\mathbf{e}=\{\Omega_{t k_{x}},\Omega_{t k_{y}},\Omega_{t \omega}\}$ and $\mathbf{b}=\{-\Omega_{\omega k_{y}},\Omega_{\omega k_{x}},\Omega_{k_{x} k_{y}}\}$. The Berry curvature in real space corresponds to the direct electromagnetic field \cite{PhysRevB.59.14915}, while the Berry curvature in reciprocal space corresponds to the reciprocal electromagnetic field \cite{PhysRevB.53.7010,PhysRevLett.97.216601,PhysRevB.110.075139}. In our case, the relevant parameter space involves a mix of time coordinates and reciprocal space coordinates, as in \cite{PhysRevB.98.064310,PhysRevResearch.7.023085}. In this non-standard system, the frequency domain Berry curvature allows us to construct a four-dimensional set of Maxwell equations with two spatial dimensions.

\textit{\textbf{Discussion}---}
Our theoretical framework can also include spatially inhomogeneity or even combined space-time inhomogeneous situations. In such cases, the momentum of the wave packet is no longer conserved, and the Berry curvature in the spatial domain also enters the equations of motion. Due to the non-standard nature of the eigenvalue equation, each term must also include a correction involving the derivative with respect to frequency, as described by the full set of {EOM} under spacetime perturbations \cite{SM}. Unlike pure time refraction, in the presence of spatial inhomogeneity, since the momentum is not conserved, the direction of the velocity after modulation is no longer parallel to the initial velocity.

Although this work focuses mainly on the localized Berry curvature effect in parameter space, the global topological properties of non-standard eigenvalue problems can also be described within our framework using the auxiliary eigenvalue. Under the on-shell condition, only two of $\{\omega, k_x, k_y\}$ are independent, so the topology of our system can be fully described by the on-shell Berry curvature defined as 
\begin{equation}
    \Omega^{on}_{\mathbf{k} \mathbf{k}}\equiv\Omega_{\mathbf{k} \mathbf{k}}+\mathbf{v}_{g}\cdot\Omega_{\mathbf{k} \mathbf{k}}+\Omega_{\mathbf{k} \mathbf{k}}\cdot\mathbf{v}_{g}.
\end{equation}
This expression takes into account the indirect dependence on $\omega$ when differentiating with respect to $\mathbf{k}$. Using the on-shell Berry curvature, the EOM has the same form as that for Bloch electrons.
It is an expression of $\omega$, $k_{x}$ and $k_{y}$. The corresponding Chern number $C^{on}=\int dk_{x} dk_{y} \Omega^{on}_{k_{x} k_{y}}|_{\lambda=0} $ should take $\omega$ as its on-shell value determined by Eq.~\eqref{onshell_condition} when performing the integration. For the two TE bands with lower and higher frequencies, the on-shell Chern numbers are $+1$ and $-1$, respectively.

The theory in this work is focused on adiabatic situations and cannot be applied to rapidly changing ones. Although how to extend the theory to non-adiabatic cases is also an important issue, there is rich physics even in adiabatic cases, as  discussed in this work, which provides guidance to relevant experiments.
Experimentally, time-varying
frequency-dispersive materials have recently been investigated under a rapidly changing mechanism \cite{wang_expanding_2025,PhysRevLett.134.183801,PhysRevB.108.104303}. Future experimental studies could explore adiabatic temporal modulation in systems with broken time-reversal or spatial-inversion symmetry, wherein the underlying physics can be described by Berry curvatures.

\textit{\textbf{Acknowledgement}---}
This work was supported by the National Key R$\&$D Program of China (2023YFA1407001) and National Natural Science Foundation of China (12234017).
\nocite{*}

\bibliography{bibliography}

\end{document}


\title{Supplemental Material for ``Frequency Domain Berry Curvature Effect on Time Refraction"}
    \author{Shiyue Deng}
    \affiliation{International Center for Quantum Design of Functional Materials (ICQD), Hefei National Research Center for Interdisciplinary Sciences at the Microscale, University of Science and Technology of China, Hefei, Anhui 230026, China}
    \affiliation{Hefei National Laboratory, University of Science and Technology of China, Hefei 230088, China}

    \author{Yang Gao}
    \affiliation{International Center for Quantum Design of Functional Materials (ICQD), Hefei National Research Center for Interdisciplinary Sciences at the Microscale, University of Science and Technology of China, Hefei, Anhui 230026, China}
    \affiliation{Hefei National Laboratory, University of Science and Technology of China, Hefei 230088, China}
    \affiliation{School of Physics, University of Science and Technology of China, Hefei, Anhui, 230026, China}

    \author{Qian Niu}
    \affiliation{International Center for Quantum Design of Functional Materials (ICQD), Hefei National Research Center for Interdisciplinary Sciences at the Microscale, University of Science and Technology of China, Hefei, Anhui 230026, China}
    \affiliation{School of Physics, University of Science and Technology of China, Hefei, Anhui, 230026, China}
    \date{\today}

    \maketitle
    
    \section{EOM of an Event Wave packet}
     In medium, the source-free Maxwell equations can be written in the following form:

    \begin{equation}
    \hat{\mathcal{H}}|\psi\rangle=i\partial_{t}\hat{\Theta}|\psi\rangle,
    \end{equation}
    where 
        $\hat{\mathcal{H}}=\left(\begin{array}{cc}
    0 & i \partial_{\mathbf{r}} \times \\
    -i \partial_{\mathbf{r}} \times & 0
    \end{array}\right),$
    $\hat{\Theta}=\left(\begin{array}{cc}
    \overleftrightarrow{\varepsilon} & 0 \\
    0 & \overleftrightarrow{\mu}
    \end{array}\right)$ and $\psi=(\mathbf{E},\mathbf{H})^{T}$ Considering the system is subjected to general spatiotemporal perturbations and approximately retains continuous translational symmetry in spacetime. For a localized wave packet, its center coordinates can be used as an approximation. The above equation can therefore be approximated as that of a local operator \cite{PhysRevB.59.14915}.
    \begin{equation}
    \hat{\mathcal{H}}_{c}\left(-i\partial_{\mathbf{r}}\right)|\psi\rangle=i\partial_{t}\hat{\Theta}_{c}\left(i\partial_{t},-i\partial_{\mathbf{r}},t_c,\mathbf{r}_c\right)|\psi\rangle
    \end{equation}
    
    When $\overleftrightarrow{\varepsilon}$ or $\overleftrightarrow{\mu}$ depends on frequency $\omega$ (dispersive optical systems), this equation becomes a non-standard eigenvalue equation for eigenvalue $\omega$. After substituting the plane wave wavefunction $\psi_{n}(\omega,\mathbf{k},\mathbf{r}_c,t_{c},\mathbf{r},t)=e^{-i(\omega t-\mathbf{k}\cdot\mathbf{r})}u_{n}(\omega,\mathbf{k},\mathbf{r}_c,t_{c})$, the orthogonality between $u_{n}(\omega,\mathbf{k},\mathbf{r}_c,t_{c})$ with the same wave vector but different frequencies is broken. As a result, we cannot use the conventional wave packet dynamics for Blcoh electrons in \cite{PhysRevB.59.14915} to analyze particle transport in these systems.
    
    To address this problem, we first convert this equation into an off-shell eigenvalue equation with respect to an auxiliary variable $\lambda$, as in \cite{dong2021semiclassical}:
    \begin{equation}
    \label{stationary_eigen_eq}
    \hat{\mathcal{L}}_{c}\left|\psi_{n}(\omega,\mathbf{k},t_c,\mathbf{r}_c)\right\rangle=\lambda_{n}\hat{\theta}_{c}\left|\psi_{n}(\omega,\mathbf{k},t_c,\mathbf{r}_c)\right\rangle,
    \end{equation}
    where $\hat{\mathcal{L}}_{c}(\omega,\mathbf{k},t_c,\mathbf{r}_c)=\mathcal{H}_{c}-\omega \hat{\Theta}_{c}$ ($\omega$ and $i\partial_{t}$ are equivalent here, as are $\mathbf{k}$ and $-i\partial_{\mathbf{r}}$). The dependence of $\hat{\mathcal{L}}_{c}$ on the adiabatic parameters $t_c$ and $\mathbf{r}_c$ represents slow perturbations of the system in spacetime, as described in \cite{PhysRevB.59.14915}, which are not the Fourier conjugates of $\omega$ and $\mathbf{k}$ included in the bra-ket notation of $\psi$. The subscript n is used to label different bands when $\hat{\mathcal{L}}$ or $\hat{\theta}$ has a matrix structure.
    
    The reason we call this an `` off-shell '' equation is that, in general, the solutions of this equation differ from those of the original equation. Only when the on-shell condition $\lambda=0$ is satisfied does it coincide with the original equation.
    
    $\hat{\theta}_{c}(\omega,\mathbf{k},t_c,\mathbf{r}_c)$ is a deliberately chosen semi-positive definite operator. Thus, the orthonormality condition for different off-shell eigenstates can be written in the following form:
    \begin{equation}
    \label{orthonormality}
        \begin{aligned}
            &\left\langle\psi_m(\omega, \mathbf{k},t_c,\mathbf{r}_c)\right| \hat{\theta}^{\frac{1}{2}}_{c}(\omega,\mathbf{k},t_c,\mathbf{r}_c)\cdot\hat{\theta}^{\frac{1}{2}}_{c}(\omega^{\prime},\mathbf{k}^{\prime},t_c,\mathbf{r}_c)\left|\psi_n\left(\omega^{\prime}, \mathbf{k}^{\prime},t_c,\mathbf{r}_c\right)\right\rangle\\
            =&\frac{1}{(2 \pi)^{d+1}} \int d t d \mathbf{r} e^{i(\omega t-\mathbf{k} \cdot \mathbf{r})} \left\langle u_m(\omega,\mathbf{k},t_c,\mathbf{r}_c)\right| \hat{\theta}^{\frac{1}{2}}_{c}(\omega,\mathbf{k},t_c,\mathbf{r}_c)\cdot\hat{\theta}^{\frac{1}{2}}_{c}(\omega^{\prime},\mathbf{k}^{\prime},t_c,\mathbf{r}_c) \left|u_{n}\left(\omega^{\prime}, \mathbf{k}^{\prime},t_c,\mathbf{r}_c\right)\right\rangle e^{-i\left(\omega^{\prime} t-\mathbf{k}^{\prime} \cdot \mathbf{r}\right)}\\
            =&\delta\left(\omega-\omega^{\prime}\right) \delta^d\left(\mathbf{k}-\mathbf{k}^{\prime}\right)\left\langle u_m(\omega,\mathbf{k},t_c,\mathbf{r}_c)\right| \hat{\theta}_{c}(\omega,\mathbf{k},t_c,\mathbf{r}_c)\left|u_{n}\left(\omega^{\prime}, \mathbf{k}^{\prime},t_c,\mathbf{r}_c\right)\right\rangle\\
            =&\delta\left(\omega-\omega^{\prime}\right) \delta^d\left(\mathbf{k}-\mathbf{k}^{\prime}\right) \delta_{m n}.
        \end{aligned}
    \end{equation}
    The inner product is defined over both time and space.

    To study the dynamics corresponding to Eq.~\eqref{stationary_eigen_eq}, we introduce an auxiliary time $\tau$, which is conjugate to the auxiliary variable $\lambda$, to parameterize the motion. Assume the wave function satisfies the following `` time ''-dependent evolution equation
    \begin{equation} 
    \label{time-dependent_eq}\hat{\mathcal{L}}_{c}\left|\Psi_{n}(\omega,\mathbf{k},t_{c},\mathbf{r}_{c},\tau)\right\rangle=i\partial_{\tau}\hat{\theta}_{c}\left|\Psi_{n}(\omega,\mathbf{k},t_{c},\mathbf{r}_{c},\tau)\right\rangle,
    \end{equation}
    Hence, the off-shell state acquires a phase as it evolves in the auxiliary time $\tau$ (note that when $\lambda=0$, the on-shell state does not evolve with respect to the auxiliary time):
    \begin{equation}
        |\Psi(\omega,\mathbf{k},\tau)\rangle=e^{-i\lambda \tau} |\psi(\omega,\mathbf{k})\rangle.
    \end{equation}
    Next, we can superimpose the eigenstates of Eq.~\eqref{time-dependent_eq} to form a wave packet that is localized in frequency, momentum, and space-time point $\{\omega_c,\mathbf{k}_c,t_c,\mathbf{r}_c\}$:
    \begin{equation}
        \label{wave_packet}
        \left|W_{n}\left(\omega_c, \boldsymbol{k}_c,t_c, \boldsymbol{r}_c,\tau\right)\right\rangle=\int d \omega d \boldsymbol{k} a\left(\omega, \mathbf{k},\tau\right)  \hat{\theta}^{\frac{1}{2}}_{c}\left(\omega, \boldsymbol{k}, t_c, \boldsymbol{r}_c\right)\left|\psi_{n}\left(\omega, \boldsymbol{k}, t_c, \boldsymbol{r}_c\right)\right\rangle.
    \end{equation}
    Here, the amplitude satisfies the normalization condition
    \begin{equation}
        \int d \omega d \boldsymbol{k} \left|a\left(\omega, \mathbf{k},\tau\right) \right|^{2}=\left\langle W | W \right\rangle =1
    \end{equation}
    and it is narrowly distributed around the center (relative to the central wavevector $\mathbf{k}_c$  and frequency $\omega_c$).
    
    For the same reason as in \cite{PhysRevB.59.14915}, we can construct the Lagrangian of the space-time wave packet expressed by Eq.~\eqref{wave_packet} in the following form based on the auxiliary time-dependent variational method
    \begin{equation}
    \label{lagrangian1}
        L_{\tau}=\left\langle W \right|i\frac{d}{d \tau}-\hat{\mathcal{L}}\left| W \right\rangle,
    \end{equation}
    with the band index omitted.

    The center coordinates of the wave packet in spacetime must satisfy:
    \begin{equation}
        \left\langle W \right|\hat{x}^{\mu}\left| W \right\rangle=x^{\mu}_{c},
    \end{equation}
    with the coordinate operator $\hat{x}_{\mu}$ corresponding to $-i\partial_{k_{\mu}}$ in 4-momentum space. $x^{\mu}=\{t,\mathbf{r}\}^{T}$, $k_{\mu}=\{-\omega,\mathbf{k}\}^{T}$, with the spacetime metric Minkowski one $\eta^{\mu \nu}=diag\{-1,1,\dots\}$. 

    We can approximate and simplify the Lagrangian into an expression involving the wave packet center in the same way as in \cite{PhysRevB.59.14915}, the result is:
    \begin{equation}
    \label{lagrangian2}
        L_{\tau}=-\lambda_{tot}+k_{c \mu}  \overset{\circ}{x_c^{\mu}}
        +\overset{\circ}{k_c^{\mu}} \left\langle v \left|\, i \frac{\partial v}{\partial k_c^{\mu}}\right.\right\rangle+\overset{\circ}{x_c^{\mu}}\left\langle v \left|\, i \frac{\partial v}{\partial x_c^{\mu}}\right.\right\rangle.
    \end{equation}
    Where $|v\rangle=\hat{\theta}^{\frac{1}{2}}|u\rangle$ and should be normalized as $\langle v|v\rangle$=1.
    A circle above denotes the derivative with respect to the auxiliary time $\tau$. Note that this differs from the solid dot, which indicates the derivative with respect to the real time $t$, i.e., $\overset{\circ}{\zeta}\equiv \frac{d \zeta}{d \tau}$. $\lambda_{tot}=\lambda_{c}+\Delta \lambda$ with
    \begin{equation}
    \hat{\mathcal{L}}_{c}\left|\psi(k^{\mu},x^{\mu}_{c})\right\rangle=\lambda_{c} \hat{\theta}_{c}\left|\psi(k^{\mu},x^{\mu}_{c})\right\rangle,
    \end{equation}
    and
    \begin{equation}
        \Delta\lambda=-\operatorname{Im}\left[\left\langle\frac{\partial v}{\partial x_c^{\mu}}\right| \left(\lambda_c-\hat{\tilde{\mathcal{L}}}_c\right)\left|\frac{\partial v}{\partial k_{\mu}}\right\rangle\right],
    \end{equation}
    where $\hat{\tilde{\mathcal{L}}}_{c}=\hat{\theta}^{-\frac{1}{2}}_{c} \hat{\mathcal{L}_{c}} \hat{\theta}^{-\frac{1}{2}}_{c}$.
    The equations of motion with respect to $\tau$ are given by the Eluer-Lagrange equation of the lagrangian \eqref{lagrangian2} as:
    \begin{subequations}
    \label{simple_lambda}
    \begin{align}
        & \overset{\circ}{t}=-\partial_\omega \lambda_{t o t}+\big(\Omega_{\omega \mathbf{r}} \cdot \overset{\circ}{\mathbf{r}}-\Omega_{\omega t} \overset{\circ}{t}\big)+\Omega_{\omega \mathbf{k}} \cdot \overset{\circ}{\mathbf{k}} \label{dtdtau} \\
        & \overset{\circ}{\mathbf{r}}=\partial_{\mathbf{k}} \lambda_{tot}-\big(\Omega_{\mathbf{k r}} \cdot \overset{\circ}{\mathbf{r}}-\Omega_{\mathbf{k} t} \overset{\circ}{t}\big)-\big(\Omega_{\mathbf{k} \mathbf{k}} \cdot \overset{\circ}{\mathbf{k}}-\Omega_{\mathbf{k} \omega} \overset{\circ}{\omega}\big)\label{drdtau} \\
        & \overset{\circ}{\omega}=\partial_t \lambda_{tot}-\Omega_{t \mathbf{r}} \cdot \overset{\circ}{\mathbf{r}}-\big(\Omega_{t \mathbf{k}} \cdot \overset{\circ}{\mathbf{k}}-\Omega_{t \omega} \overset{\circ}{\omega}\big)\label{domegadtau} \\
        & \overset{\circ}{\mathbf{k}}=-\partial_{\mathbf{r}} \lambda_{tot}+\big(\Omega_{\mathbf{r r}} \cdot \overset{\circ}{\mathbf{r}}-\Omega_{\mathbf{r} t} \overset{\circ}{t}\big)+\big(\Omega_{\mathbf{r k}} \cdot \overset{\circ}{\mathbf{k}}-\Omega_{\mathbf{r} \omega} \overset{\circ}{\omega}\big)\label{dkdtau}.
    \end{align}
    \end{subequations}
    where $\Omega_{p q}\equiv -2\operatorname{Im}\langle \partial_{p}v|\partial_{q}v\rangle$.
    To obtain the equation of motion with respect to real time t, we need to eliminate the auxiliary time $\tau$. For example, we can divide Eq.~\eqref{drdtau} by Eq.~\eqref{dtdtau} (Ensuring that the terms containing derivatives with respect to $\tau$ are moved to the same side of the equation during the division process) and
    \begin{equation}
        \begin{aligned}
    &\frac{\partial_{\mathbf{k}} \lambda_{tot}}{\partial_\omega \lambda_{tot}}=\frac{\overset{\circ}{\mathbf{r}}+\big(\Omega_{\mathbf{k r}} \cdot\overset{\circ}{\mathbf{r}}+\Omega_{\mathbf{k} t} \overset{\circ}{t}\big)+\big(\Omega_{\mathbf{k k}} \cdot \overset{\circ}{\mathbf{k}}+\Omega_{\mathbf{k} \omega} \overset{\circ}{\omega}\big)}{-\overset{\circ}{t}+\big(\Omega_{\omega \mathbf{r}}\cdot\overset{\circ}{\mathbf{r}}+\Omega_{\omega t} t\big)+\Omega_{\omega \mathbf{k}} \overset{\circ} {\mathbf{k}}}\\
    &\downarrow \text { divide both sides by }\overset{\circ}{t}\\
    &=\frac{\dot{\mathbf{r}}+\left(\Omega_{\mathbf{k r}} \cdot\dot{\mathbf{r}}+\Omega_{\mathbf{k} t}\right)+\big(\Omega_{\mathbf{k k}} \cdot \dot{\mathbf{k}}+\Omega_{\mathbf{k} \omega} \dot{\omega}\big)}{-1+\big(\Omega_{\omega \mathbf{r}}\cdot \dot{\mathbf{r}}+\Omega_{\omega t}\big)+\Omega_{\omega \mathbf{k}}\cdot \dot{\mathbf{k}}},
        \end{aligned}
    \end{equation}
    where the dot above represents the derivative of $t$. In the above derivation, we have used the relation $\overset{\circ}{\zeta}/\overset{\circ}{t}=\dot{\zeta}$. Rearrange the above equation, one can obtain the equation of motion for $\mathbf{r}$. Then, by dividing Eq.~\eqref{domegadtau} and \eqref{dkdtau} by \eqref{dtdtau}, respectively. Then using the same method, one can obtain the equations of motion for $\mathbf{k}$ and $\mathbf{\omega}$ with the derivative of t. Thus, we arrive at the final equations as shown below (up to the first order of spacetime perturbation).
    \begin{subequations}
    \label{EOM}
        \begin{align}
        \dot{\mathbf{r}}&=\mathbf{v}_{g}- \big( \left( \Omega_{\mathbf{k}\mathbf{r}} + \mathbf{v}_{g} \Omega_{\omega\mathbf{r}} \right) \cdot\dot{\mathbf{r}} + \left( \Omega_{\mathbf{k}t} + \mathbf{v}_{g} \Omega_{\omega t} \right) \big) - \big( \left( \Omega_{\mathbf{k}\mathbf{k}} + \mathbf{v}_{g} \Omega_{\omega\mathbf{k}} \right) \cdot\dot{\mathbf{k}} + \Omega_{\mathbf{k}\omega} \dot{\omega} \big); \\
        \dot{\mathbf{k}}&=\mathbf{F}+ \big( \left( \Omega_{\mathbf{r}\mathbf{r}} - \mathbf{F} \Omega_{\omega\mathbf{r}} \right)\cdot \dot{\mathbf{r}} + \left(\Omega_{\mathbf{r}t}-\mathbf{F}\Omega_{\omega t}\right) \big) + \big( \left( \Omega_{\mathbf{r}\mathbf{k}} - \mathbf{F} \Omega_{\omega\mathbf{k}} \right) \cdot \dot{\mathbf{k}} + \Omega_{\mathbf{r}\omega} \dot{\omega} \big); \\
        \dot{\omega}&=P + \left( \Omega_{t\mathbf{r}} + P \Omega_{\omega\mathbf{r}} \right) \cdot \dot{\mathbf{r}} + \left( \Omega_{t\mathbf{k}} + P \Omega_{\omega\mathbf{k}} \right) \cdot\dot{\mathbf{k}} + \Omega_{t\omega} \dot{\omega}.
        \end{align}
    \end{subequations}
    where $\mathbf{v}_{g}\equiv\partial_\mathbf{k}\lambda_{tot}/(-\partial_\omega\lambda_{tot})$, $\mathbf{F}\equiv-\partial_{\mathbf{r}}\lambda_{tot}/(-\partial_{\omega}\lambda_{tot})$ and $P\equiv\partial_{t}\lambda_{tot}/(-\partial_{\omega}\lambda_{tot})$, corresponding to group velocity, conservative force, and chirp rate, respectively.

    \section{Solutions of the Off-shell Equation}
        In this section, we will explicitly solve the off-shell Maxwell eigenvalue equations and discuss the physical implications of each solution.

        By substituting the plane wave solution $\psi(\omega,\mathbf{k},\mathbf{r},t)=e^{-i(\omega t-\mathbf{k}\cdot\mathbf{r})}u(\omega,\mathbf{k})$ into the off-shell eigenvalue equation in the main text, we can obtain the eigenvalue equation for polarization state $u=\left(\mathbf{u}_{e},\mathbf{u}_{h}\right)^{T}$.
    \begin{equation}
    \label{off-shell Maxwell eq2}
    \hat{\mathcal{L}}\left(\omega,\mathbf{k}\right)u(\omega,\mathbf{k})=\lambda\hat{\theta}(\omega)u(\omega,\mathbf{k}),
    \end{equation}
    where $\hat{\mathcal{L}}\left(\omega,\mathbf{k}\right)\equiv e^{-i\mathbf{k}\cdot\mathbf{r}}\hat{\mathcal{L}}\left(\omega,-i \partial_{\mathbf{r}}\right)e^{i\mathbf{k}\cdot\mathbf{r}}=\hat{\mathcal{H}}\left(\mathbf{k}\right)-\omega\hat{\Theta}(\omega)$.
    Since $-i \partial_{\mathbf{r}}$ only acts on the phase of the plane wave $e^{i\mathbf{k}\cdot\mathbf{r}}$ and not on $u$, we can replace $-i \partial_{\mathbf{r}}$ with $\mathbf{k}$ in $\tilde{\mathcal{L}}$. 
    In the system we are considering, the specific form of $\hat{\mathcal{L}}$, corresponding to the dielectric function defined in the main text, is:
    \begin{equation}
        \hat{\mathcal{L}}=\begin{pmatrix}
        -\omega(1-\alpha) & -i\omega\alpha\eta & 0 & 0 & 0 & -k_y\\
        i\omega\alpha\eta & -\omega(1-\alpha) & 0 & 0 & 0 & k_x\\
        0 & 0 & -\omega\varepsilon & k_y & -k_x & 0\\
        0 & 0 & k_y & -\omega & 0 & 0\\
        0 & 0 & -k_x & 0 & -\omega & 0\\
        -k_y & k_x & 0 & 0 & 0 & -\omega\\
        \end{pmatrix}
    \end{equation}
    in which $\varepsilon=1-\frac{\omega_{p}^{2}}{\omega^{2}}$, $\alpha=\frac{\omega_{p}^{2}}{\omega^{2}-\omega_{B}^{2}}$ and $\eta=\frac{\omega_{B}}{\omega}$.

    Here, we choose the operator $\hat{\theta}$ to be the energy density operator without considering the breaking of time-reversal symmetry, which takes the following form:
    \begin{equation}
        \hat{\theta}=\text{diag}\{\theta,\theta,\theta,1,1,1\}.
    \end{equation}
    where $\theta=\partial_{\omega}(\omega\varepsilon)=1+\frac{\omega_{p}^{2}}{\omega^{2}}$.
    
    As mentioned earlier, the solutions in this system can be decoupled into TE and TM modes. The equations for the TM modes (which, in this system, are simpler because they are unaffected by the breaking of time-reversal symmetry) are:
    \begin{subequations}
    \label{TM_eigen_eq}
    \begin{align}
    - \omega \varepsilon u_{ez} + k_y u_{hx} - k_x u_{hy} &= \lambda \theta u_{ez}, \\
    k_y u_{ez} - \omega u_{hx} &= \lambda u_{hx}, \\
    -k_x u_{ez} - \omega u_{hy} &= \lambda u_{hy}.
    \end{align}
    \end{subequations}
    Evidently, this set of equations has a root $\lambda_{0}=-\omega$. The wave function corresponding to this root satisfies $u_{ez}=0$ and $k_{y}u_{hx}=k_{x}u_{hy}$. This root is derived from the on-shell zero-frequency mode (where $\omega$ is also zero when taking the on-shell condition $\lambda=0$). However, the wave function corresponding to this eigenvalue does not satisfy Gauss's law for magnetism, $\mathbf{k}\cdot\mathbf{u}_{h}=0$, so it is unphysical.
    
    The other two eigenvalues of the TM mode are given by the following quadratic equation:
    \begin{equation}
        (\lambda \theta+\omega \varepsilon)(\lambda+\omega)=k^2,
    \end{equation}
    which are $\lambda_{TM\pm}=\frac1{2\theta}{\left(-\omega(\theta+\varepsilon)\pm\sqrt{\omega^2(\theta-\varepsilon)^2+4\theta k^2}\right)}$. They are derived from the on-shell positive and negative frequency solutions, as the zero solutions of these two roots, respectively, give the dispersion relations for the positive and negative frequencies $\omega=\pm\frac{k}{\sqrt{\varepsilon}}$.

\begin{figure}[t]
    \centering
    \includegraphics[width=0.8\columnwidth]{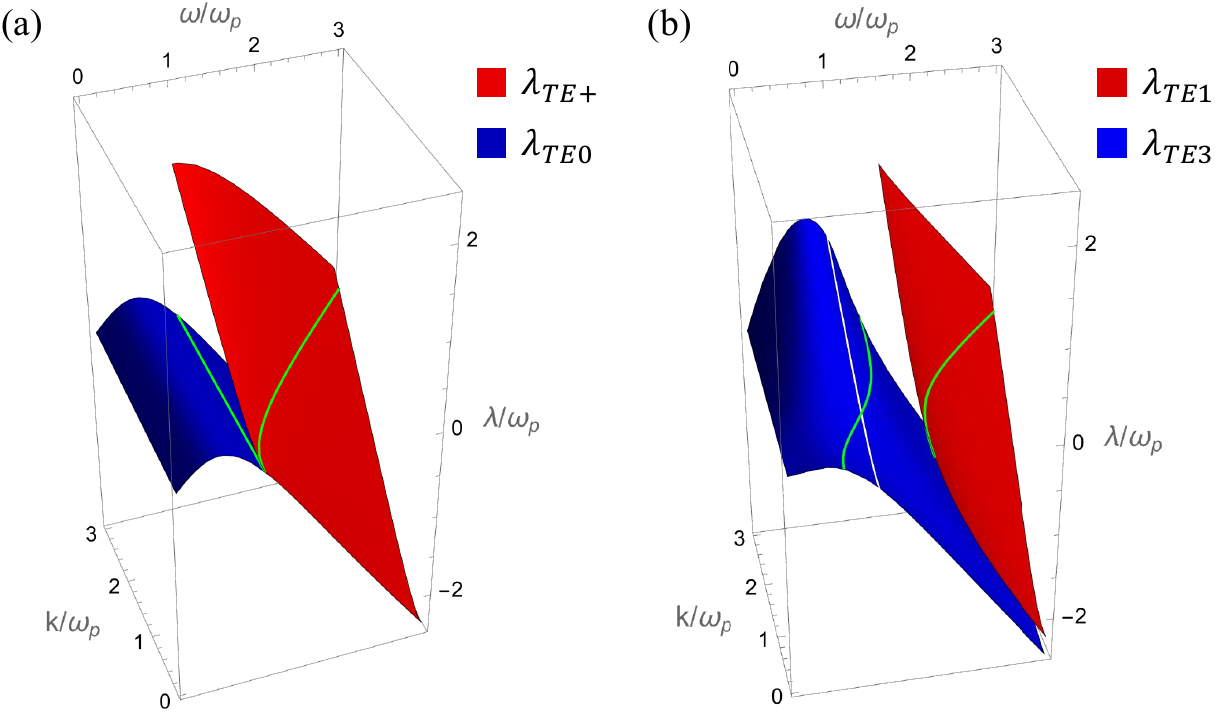}
    \caption{Dispersions of $\lambda$ with respect to $\omega$ and $k$. Green lines represent their intersection lines with the $\lambda=0$ plane. (a) At $\omega_{B}=0$, (b) at $\omega_{B}=\omega_{p}$.  For clarity, we only plot two $\lambda$ bands that contribute to on-shell dispersions for positive frequency.
    } 
    \label{lambda_bands}
    \end{figure}

    In this gyroelectric system, the time-reversal symmetry-breaking element only couples with the electric field in the x-y plane, and therefore it does not affect the eigen-equation of the TM modes. However, it will enter the eigen-equation of the TE modes, resulting in the lifting of the degeneracy between the TE and TM modes. The eigen-equations for the TE modes are as follows:

    \begin{subequations}
    \label{TE_eigen_eq}
    \begin{align}
        -\omega(1-\alpha) u_{ex} - i\omega\alpha\eta u_{ey} - k_y u_{hz} &= \lambda\theta u_{ex},\\
        i\omega\alpha\eta u_{ex} -\omega(1-\alpha)u_{ey} + k_x u_{hz} &= \lambda\theta u_{ey},\\
        -k_y u_{ex} + k_x u_{ey} - \omega u_{hz} &= \lambda u_{hz}.
    \end{align}
    \end{subequations}
    These equation also contains a root $\lambda_{0}=-\omega$, whose corresponding wave function does not satisfy Gauss's law $\mathbf{k}\cdot\mathbf{u}_{e}=0$. The other three roots are given by the following cubic equation for $\lambda$:
    \begin{equation}
    \label{TE_lambda_eigen_eq}
    (\lambda+\omega)\left(\left(\lambda\theta+\omega(1-\alpha)\right)^{2}-\omega^{2}\alpha^{2}\eta^{2}\right) = \left(\lambda\theta+\omega(1-\alpha)\right)k^2.
    \end{equation}
    When $\omega_{B}=0$ (which means $\eta=0$ and $1-\alpha=\varepsilon$), the three roots of this cubic equation are:
    \begin{subequations}
    \label{simple_lambda}
    \begin{align}
        \lambda_{TE0}&=-\frac{\varepsilon\omega}{\theta}=\frac{\omega(\omega_p^2-\omega^2)}{\omega^2+\omega_p^2};\label{lambda_TE_0}\\
        \lambda_{TE+}&=\lambda_{TM+}=\frac{\omega^2\left(-\omega+\sqrt{k^2+\frac{k^2 \omega_p^2}{\omega^2}+\frac{\omega _p^4}{\omega^2}}\right)}{\omega^2+\omega_p^2};\label{lambda_TE_plus}\\
        \lambda_{TE-}&=\lambda_{TM-}=-\frac{\omega^2\left(\omega+\sqrt{k^2+\frac{k^2 \omega_p^2}{\omega^2}+\frac{\omega _p^4}{\omega^2}}\right)}{\omega^2+\omega_p^2},\label{lambda_TE_minus}
    \end{align}
    \end{subequations}
    which are shown in Fig.~\ref{lambda_bands}(a), alone with their intersection lines with the $\lambda = 0$ plane.

    For the general case where $\omega_{B}\neq 0$, the dispersions of the three bands can be obtained from the roots of the cubic equation, which are:
    \begin{subequations}
    \label{exact_lambda}
    \begin{align}
        \lambda_{TE1}&=2 \sqrt{-\frac{p}{3}} \cos\left(\frac{\phi}{3}\right) - \frac{B}{3};\\
        \lambda_{TE2}&=2 \sqrt{-\frac{p}{3}} \cos\left(\frac{\phi}{3} + \frac{2\pi}{3}\right) - \frac{B}{3};\\
        \lambda_{TE3}&=2 \sqrt{-\frac{p}{3}} \cos\left(\frac{\phi}{3} + \frac{4\pi}{3}\right) - \frac{B}{3},
    \end{align}
    \end{subequations}
    where 
    \begin{equation}
        B = \frac{b}{a}, \quad C = \frac{c}{a}, \quad D = \frac{d}{a}, \quad
        p = C - \frac{B^2}{3}, \quad q = \frac{2 B^3}{27} - \frac{B C}{3} + D, \quad
        \phi = \arccos\left(-\frac{q}{2} \sqrt{-\frac{27}{p^3}}\right),
    \end{equation}
    and
    \begin{equation}
    \begin{aligned}
    &a = \theta^2, \quad
    b = \omega\left(\theta^{2} + 2\theta(1-\alpha)\right), \\
    &c = \omega^{2}\left((1-\alpha)^{2} - \alpha^{2}\eta^{2} + 2\theta(1-\alpha)\right) - \theta k^{2}, \quad
    d = \omega^{3}\left((1-\alpha)^{2} - \alpha^{2}\eta^{2}\right) - \omega(1-\alpha)k^{2}.       
    \end{aligned}
    \end{equation}
    One can numerically verify that when $\omega_{B}=0$, these three roots correspond respectively to the three roots given in \eqref{simple_lambda}. Their dispersions at $\omega_{B}=1$ are plotted in Fig.~\ref{lambda_bands}(b).

    Then, by substituting equation \eqref{exact_lambda} into the first two equations of \eqref{TE_eigen_eq}, one can obtain the wave functions of the three bands, which can be expressed as:
    \begin{subequations}
    \begin{align}
    u_{ex} &=  \frac{\left( i k_x \alpha \eta \omega + k_y \left( \theta \lambda + \omega - \alpha \omega \right) \right)u_{hz}}{- \theta^{2} \lambda^{2} + 2 \left(\alpha-1 \right) \theta \omega \lambda+ \left(\alpha^{2} \left(\eta^{2}-1 \right)+ 2\alpha -1 \right) \omega^{2}}, \label{uex} \\
    u_{ey} &= \frac{\left( i k_y \alpha \eta \omega - k_x \left( \theta \lambda + \omega - \alpha \omega \right) \right)u_{hz}}{- \theta^{2} \lambda^{2} + 2 \left(\alpha-1 \right) \theta \omega \lambda+ \left(\alpha^{2} \left(\eta^{2}-1 \right)+ 2\alpha -1 \right) \omega^{2}}. \label{uey}
    \end{align}
    \end{subequations}
    
    After normalizing $u$ with $u^{\dagger}\cdot \hat{\theta}\cdot u=1$, one can eliminate the variable $u_{hz}$ and obtain the normalized wave functions as functions of $k_x$, $k_y$, $\omega$ and $\omega_{p}$, and then calculate the Berry curvatures $\Omega_{pq}=-2\operatorname{Im}\partial_p{\left(\hat{\theta}_c^{\frac{1}{2}} u\right)}^\dagger\cdot\partial_q\left(\hat{\theta}_c^{\frac{1}{2}} u\right)$ and gradient correction of dispersion $\Delta\lambda=-\operatorname{Im}\partial_{x^\mu}\left(\hat{\theta}_c^{\frac{1}{2}} u\right)^\dagger\cdot(\lambda_c-\hat{\tilde{\mathcal{L}_c}})\cdot\partial_{k_\mu}\left(\hat{\theta}_c^{\frac{1}{2}} u\right)$ in the equations of motion. We present the final normalized wave function below.

    \begin{subequations}
    \label{nomalized_wave_function}
    \begin{align}
    u_{ex} &= \frac{- i k_{x} \alpha \eta \omega - k_{y} \left( \theta \lambda + \omega - \alpha \omega \right)}{\left( \theta \lambda + \omega + \alpha \left( \eta -1 \right) \omega \right)\left( \theta \lambda - \left( \alpha + \alpha \eta -1 \right) \omega \right)\sqrt{1 + \frac{1}{2} k^{2} \theta\left(\frac{1}{\left( \theta \lambda + \omega + \alpha \left(\eta -1 \right) \omega \right)^{2}}+ \frac{1}{\left( \theta \lambda - \left( \alpha + \alpha \eta -1 \right) \omega \right)^{2}}\right)}};\\
    u_{ey} &= \frac{- i k_{y} \alpha \eta \omega + k_{x} \left( \theta \lambda + \omega - \alpha \omega \right)}{\left( \theta \lambda + \omega + \alpha \left(\eta -1 \right) \omega \right)\left( \theta \lambda - \left( \alpha + \alpha \eta -1 \right) \omega \right)\sqrt{1 + \frac{1}{2} k^{2} \theta\left(\frac{1}{\left( \theta \lambda + \omega + \alpha \left( \eta -1 \right) \omega \right)^{2}}+ \frac{1}{\left( \theta \lambda - \left( \alpha + \alpha \eta -1 \right) \omega \right)^{2}}\right)}}; \\
    u_{hz} &= \frac{1}{\sqrt{1 + \frac{1}{2} k^{2} \theta\left(\frac{1}{\left( \theta \lambda + \omega + \alpha \left( \eta - 1 \right) \omega \right)^{2}}+ \frac{1}{\left( \theta \lambda - \left(  \alpha + \alpha \eta -1 \right) \omega \right)^{2}}\right)}}.
    \end{align}
    \end{subequations}
    When calculating the Berry curvature, $\omega$, $\mathbf{k}$ are independent and $\lambda$ can go off 0 shell. Firstly, we need to differentiate the wave function containing $\lambda$ with respect to various parameters, e.g., $\partial_{\omega}u=\partial_{\omega}u(\omega,\mathbf{k},\lambda(\omega,\mathbf{k}))+\frac{\partial \lambda}{\partial \omega}\partial_{\lambda}u(\omega,\mathbf{k},\lambda(\omega,\mathbf{k}))$, and then set $\lambda=0$ (on-shell).

    \section{Proof of zero $\Omega_{\omega t}$}
    In this section, we give a simple proof of $\Omega_{\omega t}=0$ for various temporal modulations. The relevant matrix elements of the TE modes are the three-by-three matrix as below (for simplicity, we set $\mathbf{k}=k\hat{y}$)
    \begin{equation}
    \hat{\mathcal{L}}_{TE}=
    \begin{pmatrix}
        -\omega(1-\alpha) & -i\omega\alpha\eta  & -k\\
        i\omega\alpha\eta & -\omega(1-\alpha) & 0\\
        -k & 0  & -\omega\\
        \end{pmatrix}.
    \end{equation}
    We actually solve the eigenvalue equation $\hat{\tilde{\mathcal{L}}}_{TE}|v_{TE}\rangle=\lambda_{TE}|v_{TE}\rangle$, with $\hat{\tilde{\mathcal{L}}}_{TE}=\hat{\theta}_{TE}^{-\frac{1}{2}}\hat{\mathcal{L}}_{TE}\hat{\theta}_{TE}^{-\frac{1}{2}}$, where $\hat{\theta}_{TE}=\text{diag}\{\theta,\theta,1\}$. $\hat{\tilde{\mathcal{L}}}_{TE}$ has the structure as follows:
    \begin{equation}
    \hat{\tilde{\mathcal{L}}}_{TE}=
    \begin{pmatrix}
        \mathcal{A} & -i\mathcal{C}  & -\mathcal{D}\\
        i\mathcal{C} & \mathcal{A} & 0\\
        -\mathcal{D} & 0  & -\mathcal{B}\\
    \end{pmatrix}.
    \end{equation}
    with $\mathcal{A},\mathcal{B},\mathcal{C},\mathcal{D}$ all real. This matrix can be transformed into a real symmetric matrix $\hat{\mathcal{L}^{\prime}}_{TE}=S \hat{\tilde{\mathcal{L}}}_{TE}S^{\dagger}$ by a constant unitary matrix $S=diag\{1,i,1\}$. Since $\hat{\mathcal{L}^{\prime}}_{TE}$ is real symmetric, its eigenfunction $|v_{TE}^{\prime}\rangle=S|v_{TE}\rangle=(v_{ex}^{\prime},v_{ey}^{\prime},v_{hz}^{\prime})=(v_{ex},iv_{ey},v_{hz})$ is a real vector. Then the Berry curvature 
    \begin{equation}
    \label{zeroBC}
    \begin{aligned}
    \Omega_{pq} &= 2\,\mathrm{Im}  \langle \partial_q v | \partial_p v \rangle  \\
    &= 2\, \frac{\partial (P, Q)}{\partial (p, q)} 
    \, \mathrm{Im} \langle \partial_Q v | \partial_P v \rangle \\
    &= 2\, \frac{\partial (P, Q)}{\partial (p, q)} 
    \, \mathrm{Im} \Bigl( 
        (\partial_Q v_{e_x})(\partial_P v_{e_x})
        + (\partial_Q v_{e_y})(\partial_P v_{e_y})
        + (\partial_Q v_{h_z})(\partial_P v_{h_z})
    \Bigr) \\
    &= 2\, \frac{\partial (P, Q)}{\partial (p, q)} 
    \, \mathrm{Im} \Bigl((\partial_Q v'_{e_x})(\partial_P v'_{e_x})
    + (\partial_Q i v'_{e_y})(\partial_P (-i) v'_{e_y})+ (\partial_Q v'_{h_z})(\partial_P v'_{h_z})\Bigr) \\
    &= 0 .
    \end{aligned}
    \end{equation}
    where the last step takes the imaginary part of a real number, so it is zero.
    $\{P,Q\}\in\{\mathcal{A},\mathcal{B},\mathcal{C},\mathcal{D}\}$, $\{p,q\}\in\{k,\omega,\omega_{p},\omega_{B}\}$ and $\frac{\partial(P,Q)}{\partial(p,q)}$ is the Jacobian between $\{P,Q\}$ and $\{p,q\}$. So, when the temporal modulation enters through $\omega_{p}$, the Berry curvature $\Omega_{\omega t}=\dot{\omega}_{p}\Omega_{\omega \omega_{p}}=0$ (same for the modulation of $\omega_{B}$). Note that we set $k_{x}=0$ and $k_{y}=k$, Eq.~\eqref{zeroBC} actually proves that $\Omega_{\omega k_{y}}(k_{x}=0)=0$ and $\Omega_{k_{y} t}(k_{x}=0)=0$, which is equivalent to $\Omega_{\omega \mathbf{k}}\perp\mathbf{k}$ and $\Omega_{\mathbf{k} t}\perp\mathbf{k}$ shown in Fig.~2 of the main text.

    \section{Dipolar correction for energy density}
    In the main text, we give the general expression for dipolar correction:
    \begin{equation}
    \label{dipolar}
        \mathbf{r}^{{\theta}^{\prime}}_{c}-\mathbf{r}^{\theta}_{c}=\frac{\left\langle\frac{1}{2}\left((\hat{r}-\mathbf{r}^{\theta}_{c})\hat{\theta}^{\prime}+\hat{\theta}^{\prime}(\hat{r}-\mathbf{r}^{\theta}_{c})\right)\right\rangle}{\langle\hat{\theta}^{\prime}\rangle}.
    \end{equation}
    For the total energy density operator $\hat{\theta}^{\prime}=\partial_{\omega}(\omega\hat{\Theta})=-\partial_{\omega}\hat{\mathcal{L}}$ the denominator is
    \begin{equation}
    \begin{aligned}
        \langle\hat{\theta}^{\prime}\rangle&=\langle W|\hat{\theta}^{-\frac{1}{2}}(-\partial_{\omega}\hat{\mathcal{L}})\hat{\theta}^{-\frac{1}{2}}|W\rangle\\
        &\approx -\langle u|\partial_{\omega}\hat{\mathcal{L}}|u\rangle\\
        &=-\langle u|\partial_{\omega}(\hat{\mathcal{L}}|u\rangle)+\langle u|\hat{\mathcal{L}}|\partial_{\omega}u\rangle\\
        &=-\langle u|\partial_{\omega}(\lambda\hat{\theta}|u\rangle) +\cancel{\lambda \langle u|\hat{\theta}|\partial_{\omega}u\rangle}\\
        &=-\partial_{\omega}\lambda \langle u|\hat{\theta}|u\rangle-\cancel{\lambda\langle u|\partial_{\omega}\hat{\theta}|u\rangle}=-\partial_{\omega}\lambda.
    \end{aligned}
    \end{equation}
    $\cancel{x}$ means those terms are zero under on-shell condition $\lambda=0$. The numerator is
    \begin{equation}
    \begin{aligned}
        \left\langle\frac{1}{2}\left((\hat{r}-\mathbf{r}^{\theta}_{c})\hat{\theta}^{\prime}+\hat{\theta}^{\prime}(\hat{r}-\mathbf{r}^{\theta}_{c})\right)\right\rangle&=\langle W|\frac{1}{2}(\hat{r}-\mathbf{r}^{\theta}_{c})\hat{\theta}^{-\frac{1}{2}}(-\partial_{\omega}\hat{\mathcal{L}})\hat{\theta}^{-\frac{1}{2}}+\frac{1}{2}\hat{\theta}^{-\frac{1}{2}}(-\partial_{\omega}\hat{\mathcal{L}})\hat{\theta}^{-\frac{1}{2}}(\hat{r}-\mathbf{r}^{\theta}_{c})|W\rangle\\
        &\approx -\operatorname{Im}\langle v|\hat{\theta}^{-\frac{1}{2}}(-\partial_{\omega}\hat{\mathcal{L}})\hat{\theta}^{-\frac{1}{2}}|\partial_{\mathbf{k}}v\rangle-\mathcal{A}_{\mathbf{k}}\langle v|\hat{\theta}^{-\frac{1}{2}}(-\partial_{\omega}\hat{\mathcal{L}})\hat{\theta}^{-\frac{1}{2}}|v\rangle\\
        &=\operatorname{Im}\langle u|\partial_{\omega}\hat{\mathcal{L}}|\partial_{\mathbf{k}}u\rangle-\mathcal{A}_{\mathbf{k}}\langle u|(-\partial_{\omega}\hat{\mathcal{L}})|u\rangle\\
    \end{aligned}
    \end{equation}
    where $\mathcal{A}_{\mathbf{k}}=i\langle v|\partial_{\mathbf{k}}v\rangle$. The first part of the above expression is
    \begin{equation}
    \begin{aligned}
        \operatorname{Im}\langle u|\partial_{\omega}\hat{\mathcal{L}}|\partial_{\mathbf{k}}u\rangle&=\operatorname{Im}\partial_{\omega}(\langle u|\hat{\mathcal{L}})|\partial_{\mathbf{k}}u\rangle-\operatorname{Im}\langle \partial_{\omega}u|\hat{\mathcal{L}}|\partial_{\mathbf{k}}u\rangle\\&=
        \operatorname{Im}\partial_{\omega}(\langle u|\lambda\hat{\theta})|\partial_{\mathbf{k}}u\rangle-\operatorname{Im}\langle \partial_{\omega}u|\hat{\mathcal{L}}|\partial_{\mathbf{k}}u\rangle\\
        &=\partial_{\omega}\lambda\operatorname{Im}\langle u|\hat{\theta}|\partial_{\mathbf{k}} u\rangle+\cancel{\lambda\operatorname{Im}\partial_{\omega}(\langle u|\hat{\theta})|\partial_{\mathbf{k}}u\rangle}-\operatorname{Im}\langle \partial_{\omega}u|\hat{\mathcal{L}}|\partial_{\mathbf{k}}u\rangle\\
        &=\partial_{\omega}\lambda\operatorname{Im}\langle v|\partial_{\mathbf{k}} v\rangle-\operatorname{Im}\langle \partial_{\omega}u|\hat{\mathcal{L}}|\partial_{\mathbf{k}}u\rangle,
    \end{aligned}
    \end{equation}
    using $\operatorname{Im}\langle v|\partial_{\mathbf{k}}v\rangle=-i\langle v|\partial_{\mathbf{k}}v\rangle=-\mathcal{A}_{\mathbf{k}}$ and $-\langle u|\partial_{\omega}\hat{\mathcal{L}}|u\rangle=-\partial_{\omega}\lambda$, we can get the simplified expression in the main text for dipolar correction of the center position:
    \begin{equation}
        \mathbf{r}^{{\theta}^{\prime}}_{c}-\mathbf{r}^{\theta}_{c}=\frac{\operatorname{Im}\langle \partial_{\omega}u|\hat{\mathcal{L}}|\partial_{\mathbf{k}}u\rangle}{\partial_{\omega}\lambda}.
    \end{equation}
    \begin{figure}[H]
    \centering
    \includegraphics[width=0.8\columnwidth]{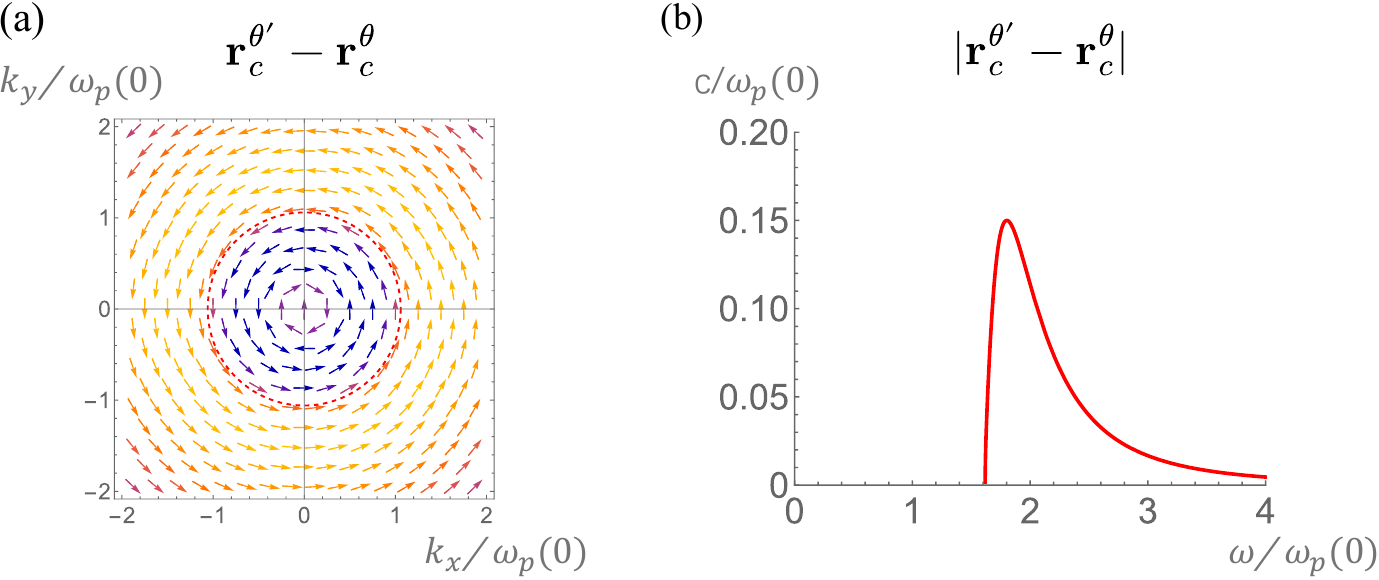}
    \caption{(a) The structure of dipolar correction \eqref{dipolar} and (b) the variation of its on-shell magnitude with frequency. The parameters are chosen the same as Fig.~2 and 3 in the main text.}
    \label{dipolar_correction}
    \end{figure}

    \section{Topology}
    The integral of $\Omega_{k_x k_y}$ with constant $\omega$ gives a quantized number. Since it is calculated by the off-shell eigenfunctions and is unphysical, we call it off-shell Chern number.
    \subsection{Off-shell Chern number}
    \begin{figure}[H]
    \centering
    \includegraphics[width=0.6\columnwidth]{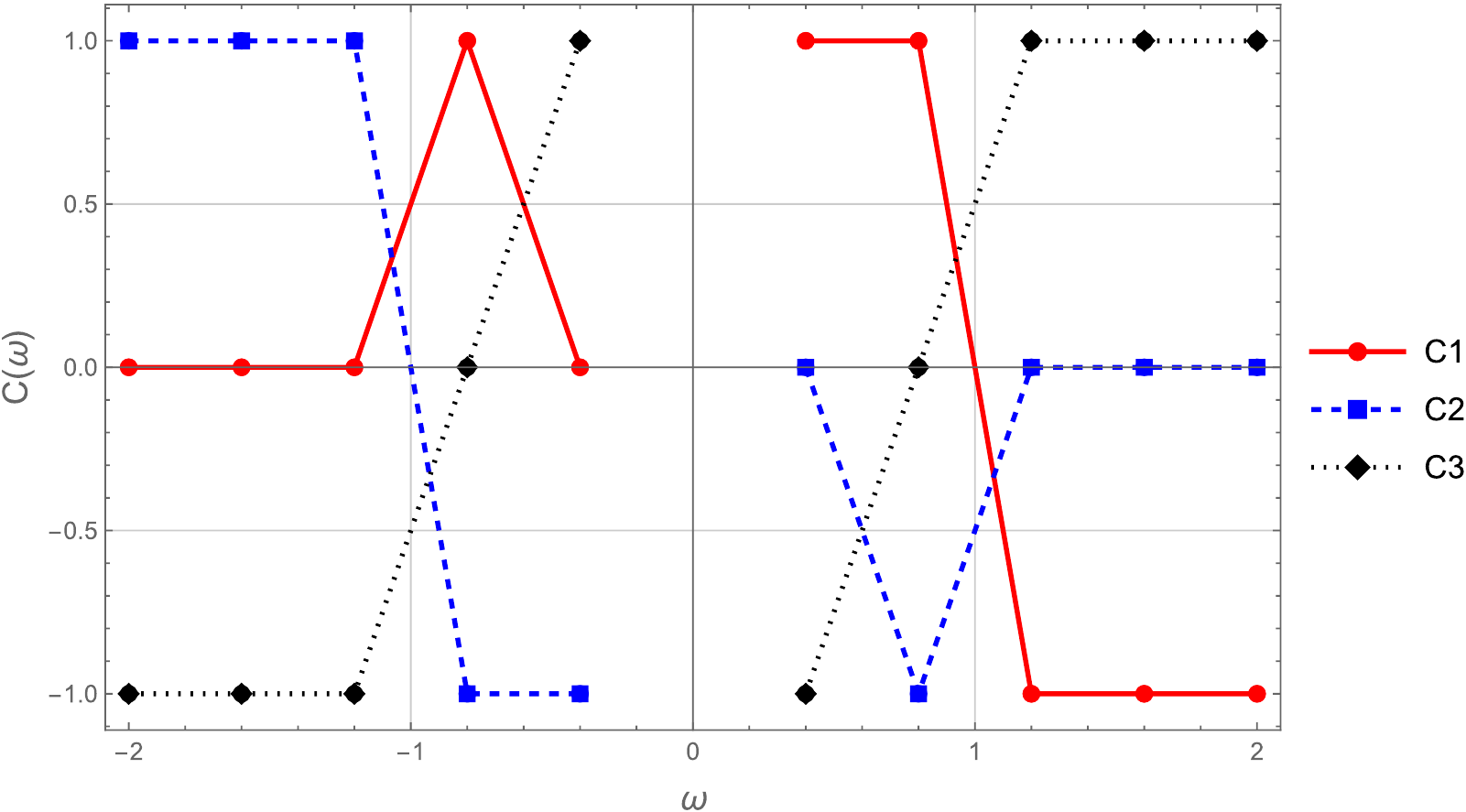}
    \caption{Off-shell Chern number $C(\omega)$ of the 3 TE bands.} 
    \label{experiment}
    \end{figure}
    The Chern numbers of the three bands sum to zero and are centrosymmetric about the point $(0,0)$.
    \subsection{On-shell Chern number}
    By defining on-shell Berry curvatures
    \begin{subequations}
    \label{onshell_Berry_curvatures}
        \begin{align}
            & \Omega_{\mathbf{kr}}^{o n}=\Omega_{\mathbf{kr}}+\frac{\partial \omega}{\partial \mathbf{k}} \Omega_{\omega \mathbf{r}}+\Omega_{\mathbf{k} \omega} \frac{\partial \omega}{\partial \mathbf{r}} \\
            & \Omega_{\mathbf{kk}}^{o n}=\Omega_{\mathbf{kk}}+\frac{\partial \omega}{\partial \mathbf{k}} \Omega_{\omega \mathbf{k}}+\Omega_{\mathbf{k} \omega} \frac{\partial \omega}{\partial \mathbf{k}} \\
            & \Omega_{\mathbf{rr}}^{o n}=\Omega_{\mathbf{rr}}+\frac{\partial \omega}{\partial \mathbf{r}} \Omega_{\omega \mathbf{r}}+\Omega_{\mathbf{r} \omega} \frac{\partial \omega}{\partial \mathbf{r}} \\
            & \Omega_{\mathbf{k} t}^{o n}=\Omega_{\mathbf{k} t}+\frac{\partial \omega}{\partial \mathbf{k}} \Omega_{\omega t}+\Omega_{\mathbf{k} \omega} \frac{\partial \omega}{\partial t}
        \end{align}
    \end{subequations}
    and using the relation $\dot{\omega}=\frac{\partial \omega}{\partial \mathbf{r}} \cdot \dot{\mathbf{r}}+\frac{\partial \omega}{\partial \mathbf{k}} \cdot \dot{\mathbf{k}}+\frac{\partial \omega}{\partial t}$, we can express Eq.~\eqref{EOM} as below
    \begin{subequations}
        \begin{align}
    \dot{\mathbf{r}} & =\frac{\partial \omega}{\partial \mathbf{k}}-\left(\Omega_{\mathbf{kr}}^{o n} \cdot \dot{\mathbf{r}}+\Omega_{\mathbf{kk}}^{o n} \cdot \dot{\mathbf{k}}\right)+\Omega_{t \mathbf{k}}^{o n} \\
    \dot{\mathbf{k}} & =-\frac{\partial \omega}{\partial \mathbf{r}}+\left(\Omega_{\mathbf{rr}}^{o n} \cdot \dot{\mathbf{r}}+\Omega_{\mathbf{rk}}^{o n} \cdot \dot{\mathbf{k}}\right)-\Omega_{t \mathbf{r}}^{o n},
        \end{align}
    \end{subequations}
    which have the same form as EOM of Bloch electrons in \cite{PhysRevB.59.14915}.
    This indicates the real topology of the system should be described by the integral of on-shell Berry curvature $\Omega^{on}_{\mathbf{kk}}$:
    \begin{equation}
        C^{on}=\int dk_{x} dk_{y} \Omega^{on}_{k_{x} k_{y}}|_{\lambda=0}.
    \end{equation}
    as mentioned in the main text. The configuration of $\Omega^{on}_{\mathbf{kk}}$ is shown in Fig.~\ref{on-shell_BC} for the lower and upper positive frequency bands of TE modes:
    \begin{figure}[!htbp]
    \centering
    \includegraphics[width=\columnwidth]{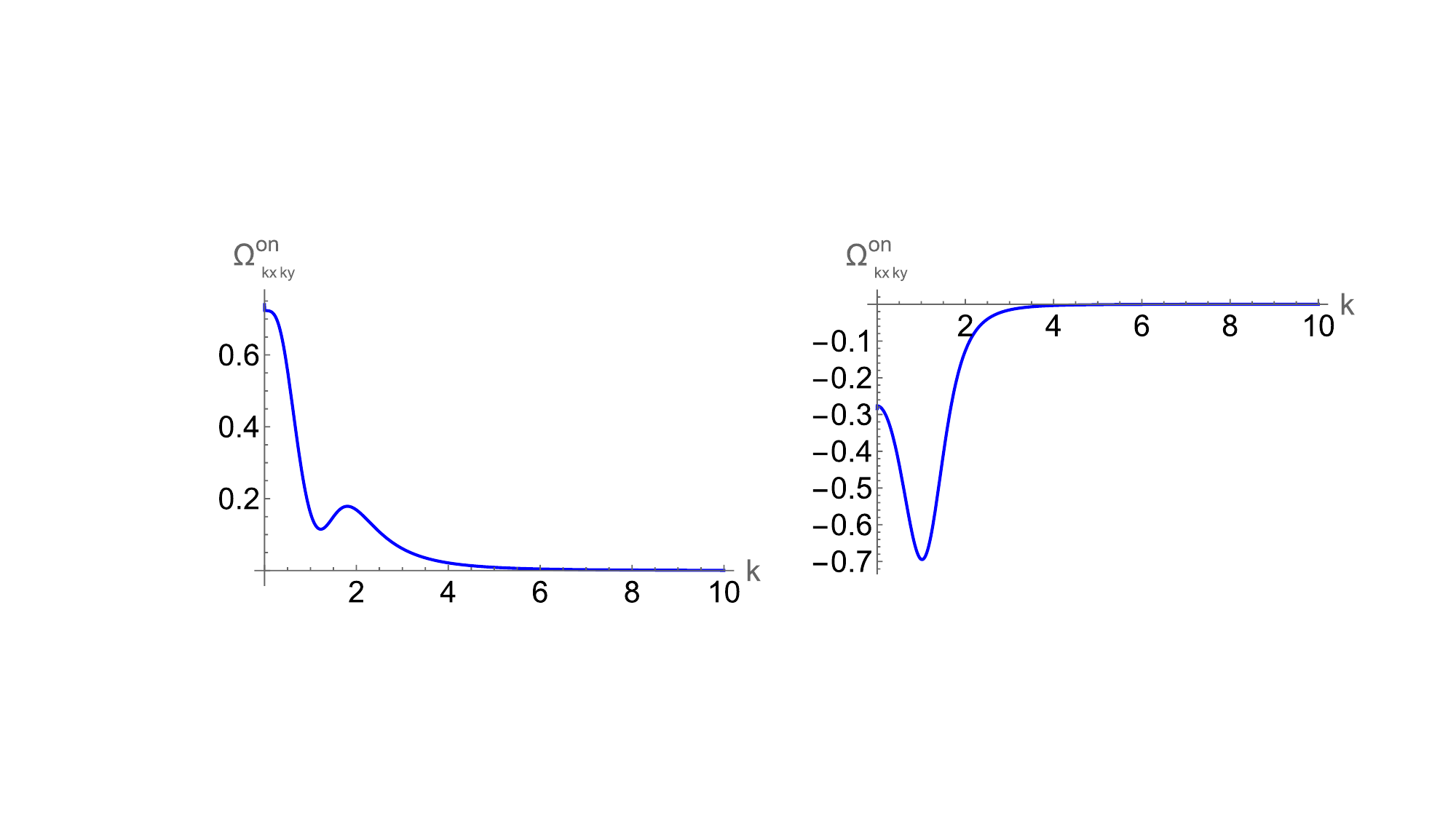}
    \caption{On-shell Berry curvature $\Omega_{k_x k_y}^{on}$ for lower (left) and upper (right) positive frequency TE bands as a function of $k$.}
    \label{on-shell_BC}
    \end{figure}
    Without the two terms $\frac{\partial \omega}{\partial k_x}\Omega_{\omega k_y}$  and $\Omega_{k_x \omega}\frac{\partial \omega}{\partial k_y}$, the calculated integral is not quantized (0.849105 for the lower band and -0.616657 for the upper band). Once these two terms are included, the result becomes quantized (+1 for the lower band and -1 for the upper band).

    \section{Experimental realization}
    As illustrated in Fig.~\ref{experiment}, the proposed proof‐of‐principle experiment can be implemented relatively easy. A uniform, static magnetic field $B_{0}$ is imposed throughout the bulk of a metallic sample, and an electromagnetic pulse of initial frequency $\omega$ and momentum $\mathbf{k}$ is emitted within the plane orthogonal to $\mathbf{B}_{0}$. At time $t=0$, the bulk plasmon frequency $\omega_{p}$ is modulated in time under the adiabatic constraint $\dot{\omega}_{p}\ll\omega_{p}^{2}$. This modulation continues until $t = t_{f}$, at which point $\omega_{p}$ is held constant again; subsequently, the velocity of the pulse realigns with its momentum direction. Throughout this process, the net displacement of the pulse transverse to $\mathbf{k}$ is given by Eq. (7) of the main text, and the final pulse frequency $\omega^{\prime}$ is obtained by integrating the corresponding equation of motion in the main text.
    \begin{figure}[H]
    \centering
    \includegraphics[width=0.6\columnwidth]{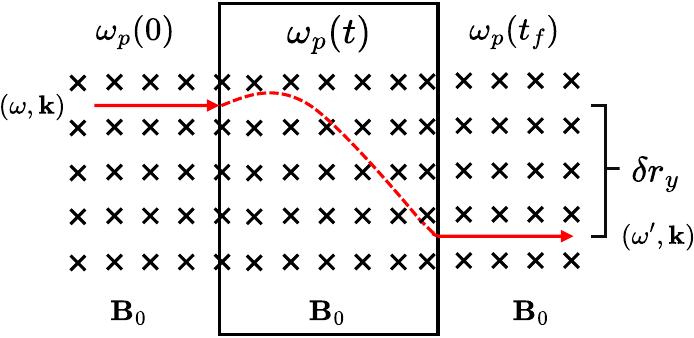}
    \caption{Schematic diagram of the experimental setup.} 
    \label{experiment}
    \end{figure}
    
    \nocite{*}
    \bibliography{supp_bibliography}